\documentclass[prd,aps,a4paper,twocolumn,eqsecnum,nofootinbib,floatfix]{revtex4} 

\newif\ifusesec
\usesectrue  
   
\usepackage{graphicx} 
\usepackage{amsmath,amsfonts,amssymb}
\usepackage{appendix}
\usepackage{mathtools}
\usepackage[caption=false,justification=justified]{subfig}
\usepackage{xcolor}
\usepackage{comment}

\newcommand{\beq}{\begin{equation}}
\newcommand{\eeq}{\end{equation}}
\newcommand{\bea}{\begin{eqnarray}}
\newcommand{\eea}{\end{eqnarray}}

\begin{document}

\title{Scalar perturbations in extremal Reissner-Nordstr\"om geometry: new methods at work}

\author{Giorgio Di Russo$^{1}$, Davide Fioravanti$^{2}$, Francesco Olivi$^{2}$, Marco Rossi$^{3}$}

\affiliation{
$^1$School of Fundamental Physics and Mathematical Sciences, Hangzhou Institute for Advanced Study, UCAS, Hangzhou 310024, China\\
$^2$Sezione INFN di Bologna, Dipartimento di Fisica e Astronomia, Universit\`a di Bologna,
Via Irnerio 46, 40126 Bologna, Italy\\
$^3$Dipartimento di Fisica dell'Universit\`a della Calabria and INFN, Gruppo collegato di Cosenza,
Arcavacata di Rende, 87036 Cosenza, Italy
}
 
\date{\today}

\begin{abstract}
We study scalar radiation emitted by a particle moving along a circular orbit around a four-dimensional extremal Reissner-Nordstr\"om black hole. In particular, we calculate the energy fluxes carried by scalar waves to infinity and across the event horizon, as well as the scalar self-force acting on the particle, and express all these quantities as post-Newtonian expansions. Our results for the energy fluxes are in agreement with previous numerical findings in the literature. To solve the radial wave equation, we employ and test two distinct analytic approaches: the first is the so-called \textit{kink method}, a novel alternative stemming from quantum integrability and the second is the Mano-Suzuki-Takasugi method. The perfect agreement of the results highlights the potential of the kink method as a complementary tool for studying wave propagation in black-hole spacetimes.
\end{abstract}

\maketitle

\section{Introduction}
\label{Intro}
Scalar self-force (SSF) effects emerge when a scalar-charged particle moving along a prescribed trajectory in curved spacetime couples to the scalar field generated by the particle itself. This field obeys a wave equation of the d'Alembert type, sourced by a term that is singular at the particle's location. This framework provides a simplified model for the more intricate problem of gravitational perturbations produced by a small mass orbiting in a background spacetime whose geometry is affected by the particle's own presence. In the gravitational case, the interaction between the particle and its self-generated field leads to the gravitational self-force (GSF) (see, e.g., Ref. \cite{Barack:2009ux} and references therein).

Although the latter problem is of greater physical interest, the former is more tractable, despite the fact that both share similar conceptual approaches and computational tools. For this reason, scalar self-force (SSF) calculations have often served as a testing ground for the gravitational self-force (GSF) problem, allowing one to identify and address technical challenges that also arise in the more general gravitational setting. The literature on this subject is extensive. In addition to the seminal contributions that established the foundations of self-force theory in curved spacetime \cite{Mino:1996nk,Quinn:1996am,Quinn:2000wa,Barack:2001gx,Barack:2001ph,Detweiler:2002mi,Poisson:2003nc}, numerous subsequent studies have investigated self-force effects in black hole backgrounds, particularly Schwarzschild and Kerr spacetimes \cite{Burko:1999zy,Barack:2000eh,Burko:2000xx,Wiseman:2000rm,Nakano:2001kw,Burko:2001kr,Barack:2002mha,Detweiler:2002gi,Nakano:2003he,Hikida:2003pi,Diaz-Rivera:2004nim,Hikida:2004hs,Ottewill:2007mz,Haas:2007kz,Damour:2009sm,Barack:2010tm,Warburton:2010eq,Wardell:2011gb,Ottewill:2012aj,Vega:2013wxa,Isoyama:2014mja,Bini:2025bll}.

More recently, SSF techniques have been employed to investigate radial infall in Schwarzschild spacetime in the weak-field regime \cite{Bini:2026cpw, Bini:2026ova}, with the resulting predictions compared against their counterparts derived within the multipolar post-Minkowskian (MPM) formalism \cite{Bini:2026ova,DiRusso:2026fqn}. We refer the reader to Ref. \cite{Blanchet:2013haa} for a review of the MPM formalism. These methods have also been extended to higher-dimensional spacetimes, beyond four dimensions, and applied to specific solutions of five-dimensional Einstein-Maxwell theory, namely topological stars and W-solitons \cite{Bianchi:2024vmi,Bianchi:2024rod,DiRusso:2025lip,Bianchi:2025aei,Bini:2025ltr,Bianchi:2025ydq}. A general discussion of these solutions can be found in Refs. \cite{Bah:2020pdz,Dima:2025tjz}. Although these configurations are formulated in five dimensions, they are not genuinely higher-dimensional in the sense of having an additional non-compact spatial dimension, as one spatial direction is compactified. Nevertheless, SSF techniques have also been applied to the genuinely five-dimensional Schwarzschild-Tangherlini spacetime, providing a framework for investigating self-force effects in higher-dimensional black hole backgrounds \cite{Bini:2026foz}.

The self-force program has traditionally relied on the post-Newtonian (PN) approximation, which assumes weak gravitational fields and non-relativistic velocities. This is largely due to the fact that the radial perturbation equations arising in black hole perturbation theory are typically Heun-like equations, -- namely Heun Equation (HE) and its confluent limits \cite{Consoli:2022eey} --, for which no general closed-form solution is known. A straightforward PN expansion of these equations is, however, insufficient to preserve the correct boundary conditions, namely purely ingoing waves at the horizon for the In-solution and purely outgoing waves at infinity for the Up-solution. The first systematic approach to overcome this difficulty, allowing solutions of the Confluent HE (CHE) to be expanded in the PN regime while consistently enforcing the appropriate boundary conditions, was developed approximately three decades ago by Mano, Suzuki, and Takasugi \cite{Mano:1996vt} (MST) and subsequently refined and extended in \cite{Sasaki:2003xr}.

More recently, alternative analytical approaches have been developed to disentangle the vast topic. For instance, Refs.~\cite{Cipriani:2025ikx,Cipriani:2026myb} exploit the correspondence between quantum Seiberg-Witten curves and the CHE arising in black hole perturbation theory, thereby addressing the powerful techniques of instanton calculus to construct solutions within a suitable expansion scheme. A key advantage of this approach is that certain contributions to the wave functions are incorporated non-perturbatively at all PN orders, thereby improving the convergence of the resulting PN series. Using this method, we have computed the flux emitted by a particle on a circular orbit around a Schwarzschild black hole to 30PN order, substantially surpassing the previous state of the art, which was limited to 22PN order \cite{Fujita:2012cm}.

In this context, a new correspondence between supersymmetric gauge theories and quantum integrable theories was found in \cite{Fioravanti:2019vxi} and gave rise to perturbative, but also non-perturbative results in the instanton coupling. Subsequently, it was translated and applied to black hole theory in \cite{Fioravanti:2021dce}. In particular, efficient non-linear integral equations, dubbed Thermodynamic Bethe Ansatz (TBA), allow calculating quasi normal modes and many gravity factors. Besides, \cite{Fioravanti:2024dkx} elaborated further the integrability approach by unveiling the prepotential as defined uniquely within the Ordinary Differential Equation (ODE) theory. Furthermore, \cite{Fioravanti:2025bts} developed the so-called kink method to provide a systematic framework for constructing solutions to Heun-like equations, and for computing the prepotential, many gravity factors (without TBA) and Love numbers. This new methodology starts directly from the differential equation and constructs Floquet and other solutions by solving {\it exactly} the corresponding Riccati equation (for the quantum momentum) at a certain order by means of special polynomials. Although its development is evidently general for any Heun-like equation, it was given some details in \cite{Fioravanti:2025bts} for what concerns the Modified Mathieu Equation (MME), the CHE and the HE itself, providing also a MST-like solution as a consequence. Instead, in the following, the kink method will be developed specifically for the Doubly Confluent Heun Equation (DCHE).

The present work investigates scalar self-force (SSF) effects for a scalar-charged particle moving along a circular equatorial orbit in the spacetime of an extremal Reissner-Nordstr\"om (eRN) black hole \cite{Bini:2006pk,Bini:2006dp,Bini:2007zzd,Bini:2008iwy,Bini:2008zzd}. 
This problem was previously investigated in Ref. \cite{Bini:2016egn},where the extremal case is approached by specializing the resulting expressions to $Q=M$. In the present work, we adopt a completely different strategy, solving the radial perturbation equation directly in the extremal geometry, where the confluence of the two horizons leads to a different differential equation, a DCHE. This formulation allows us to impose the appropriate boundary conditions directly in the extremal spacetime and provides an independent analytical framework for computing the scalar self-force and the associated energy fluxes.

To this aim, we are going to develop and compare both the kink method and an extension of the MST method to the radial equation in eRN spacetime. These results will be shown to be in agreement with each other and eventually with the numerical findings reported in Ref. \cite{Castillo:2018ibo}. Our PN expressions reproduce the numerical values with residuals that scale as expected from the truncation order, and they satisfy the energy-balance relation $F_t=\Gamma\,(\dot E^{\infty}+\dot E^{H})$ identically. To the best of our knowledge, the results presented here are therefore the first reliable analytic PN expressions for the scalar fluxes, self-field and self-force in eRN spacetime.

The paper is organized as follows. In Sec. \ref{eRNsection}, we study the motion of massive particles in eRN spacetime, deriving the geodesic equations and analyzing circular orbits. To complete the study of geodesic motion, and in view of future research directions, we also consider hyperbolic trajectories and compute the associated scattering angle. This analysis will allow us to review a remarkable symmetry of this spacetime, originally discovered by Couch and Torrence \cite{Couch:1984tr,Cvetic:2021lss,Bianchi:2021yqs,Bianchi:2022wku,Akhond:2026suo}, which remains relatively unexplored and not fully understood.
In Sec. \ref{waveEQ}, we analyze the radial wave equation using the so-called kink method and an MST-like approach, suitably adapted to the DCHE. We show that the two approaches yield equivalent solutions, providing an independent cross-check of the analytical results. In Sec. \ref{scalar_source}, we derive the scalar source term associated with the orbiting particle, which acts as the forcing term in the scalar field equation in eRN spacetime. In Sec. \ref{Energy}, we compute the radiative energy fluxes carried by scalar radiation to infinity and to the horizon. In Sec. \ref{self_field}, we compute the scalar self-force acting on the particle and analyze its properties in eRN spacetime. In Sec. \ref{scalar_self_force}, we present our results for the scalar self-force and discuss their implications in eRN spacetime. At the end in Sec. \ref{conclusions} we report some final considerations and outlook. We will work in natural units $G=c=1$.

\section{Extremal Reissner-Nordstr\"om timelike geodesics}
\label{eRNsection}
Let us consider the eRN spacetime with
line element written in standard Schwarzschild-like coordinates $(t,r,\theta,\phi)$ as
\begin{align}
    ds^2&=-\left(1-\frac{Q}{r}\right)^2dt^2+\frac{dr^2}{\left(1-\frac{Q}{r}\right)^2}\nonumber\\
    &+r^2\left(d\theta^2+\sin^2\theta d\phi^2\right)\,.
\end{align}
The 4-velocities are written in terms of the conjugate momenta as 
\begin{align}\label{4vel}
u^{t}&=\frac{dt}{d\tau}=\frac{r^2E}{(r-Q)^2}\,,\nonumber\\
u^{r}&=\frac{dr}{d\tau}=\frac{(r-Q)^2P_r}{r^2}\,,\nonumber\\
u^{\theta}&=\frac{d\theta}{d\tau}=\frac{P_\theta}{r^2}\,,\nonumber\\
u^{\phi}&=\frac{J}{r^2\sin^2\theta}\,,
\end{align}
where $E$ and $J$ are the conserved energy and angular momentum, while the other two canonical momenta are constrained by the Hamiltonian mass-shell condition
\begin{equation}
    \mathcal{H}=g^{\mu\nu} P_\mu P_\nu=-\mu^2\,.
\end{equation}
The previous relation is solved as
\begin{align}
P_r^2\frac{(r-Q)^2}{r^2}-\frac{r^2E^2}{(r-Q)^2}+\mu^2+\frac{P_\theta^2}{r^2}+\frac{J^2}{r^2\sin^2\theta}=0\,.
\end{align}
The separation between the radial and angular motion is achieved through the introduction of the Carter constant \cite{10.1093/oso/9780198503705.001.0001}
\begin{align}\label{geoeq}
&P_\theta^2+\frac{J^2}{\sin^2\theta}=K^2\,,\nonumber\\
&P_r^2=Q_r(r,E)=\frac{r^4E^2}{(r-Q)^4}-\frac{K^2+r^2\mu^2}{(r-Q)^2}\,.
\end{align}
Being the geometry spherically symmetric, we can focus without loss of generality on the equatorial motion
\beq
\theta=\frac{\pi}{2},\quad P_\theta=0,\quad K=J\,.
\eeq
\subsection{Hyperbolic-like geodesics}
The classical turning points are defined as the inversion points of the radial velocity i.e. through the relation $Q_r(r,E)=0$. 

Although the previous condition can be solved analytically, it is convenient to introduce the impact parameter
\begin{equation}
    b=\frac{J}{\sqrt{E^2-\mu^2}}\,,
\end{equation}
and expand the roots in the large impact parameter regime. This expansion corresponds to the post-Minkowskian (PM) expansion, characterized by $Q/b\ll 1$
\begin{widetext}
\begin{align}
r_1&=b-\frac{E^2 Q}{E^2-\mu ^2}+\frac{E^2 Q^2 \left(3 \mu ^2-2
   E^2\right)}{2 \left(E^2-\mu
   ^2\right)^2b}-\frac{2 E^2 Q^3}{(E^2-\mu ^2)b^2}+\frac{E^2 Q^4 \left(-40 E^6+104
   E^4 \mu ^2-85 E^2 \mu ^4+20 \mu
   ^6\right)}{8 b^3 \left(E^2-\mu
   ^2\right)^4}+O\left(\frac{1}{b^4}\right)\,,\nonumber\\
r_2&=Q+   \frac{E Q^2}{b\sqrt{E^2-\mu ^2}}+\frac{2 E^2 Q^3}{(E^2-\mu ^2)b^2}+\frac{E Q^4 \left(10 E^2-\mu
   ^2\right)}{2 \left(E^2-\mu
   ^2\right)^{3/2}b^3}+\frac{E^2 Q^5 \left(14 E^2-3 \mu
   ^2\right)}{\left(E^2-\mu
   ^2\right)^2b^4}+O\left(\frac{1}{b^5}\right)\,,\nonumber\\
   r_3&=Q-\frac{E Q^2}{b\sqrt{E^2-\mu ^2}}+\frac{2 E^2 Q^3}{(E^2-\mu ^2)b^2}-\frac{E Q^4 \left(10 E^2-\mu
   ^2\right)}{2 \left(E^2-\mu
   ^2\right)^{3/2}b^3}+\frac{E^2 Q^5 \left(14 E^2-3 \mu
   ^2\right)}{\left(E^2-\mu
   ^2\right)^2b^4}+O\left(\frac{1}{b^5}\right)\,,\nonumber\\
   r_4&=-b-\frac{E^2 Q}{E^2-\mu ^2}-\frac{E^2 Q^2 \left(3 \mu ^2-2
   E^2\right)}{2 \left(E^2-\mu
   ^2\right)^2b}-\frac{2 E^2 Q^3}{(E^2-\mu ^2)b^2}-\frac{E^2 Q^4 \left(-40 E^6+104
   E^4 \mu ^2-85 E^2 \mu ^4+20 \mu
   ^6\right)}{8 b^3 \left(E^2-\mu
   ^2\right)^4}+O\left(\frac{1}{b^4}\right)\,,
\end{align}
\end{widetext}
where their ordering is 
\begin{equation}
    r_1>r_2>r_3>0>r_4\,.
\end{equation}
On the equatorial plane, the deflection angle is identified by 
\begin{align}
    \Delta\Phi(A,B)&=\int_A^B\frac{d\phi}{dr}dr=\int_A^B \frac{bdr}{\sqrt{\mathcal{P}_4(r,E)}}\,,\nonumber\\
    \mathcal{P}_4(r,E)&=r^4+\frac{2Q\mu^2}{E^2-\mu^2}r^3-\left(\frac{Q^2\mu^2}{E^2-\mu^2}+b^2\right)r^2\nonumber\\
    &+2b^2Qr-b^2Q^2\nonumber\\
    &=(r-r_1)(r-r_2)(r-r_3)(r-r_4)\,,
\end{align}
where we used \eqref{4vel}. The scattering angle is defined as
\begin{equation}
    \chi(A,B)=2\Delta\Phi(A,B)-\pi\,.
\end{equation}
The angle described from infinity to the minimal approach is
\begin{equation}\label{scatt}
   \Delta\Phi(\infty,r_1)=\frac{2b}{\sqrt{r_{13}r_{24}}}F\Bigg[\arcsin\sqrt{\frac{r_{24}}{r_{14}}},\frac{r_{23}r_{14}}{r_{13}r_{24}}\Bigg]\,,
\end{equation}
where $F\Big[\alpha,m\Big]$ represents the incomplete elliptic integral of the first kind
\begin{equation}
 F\Big[\alpha,m\Big]=\int_0^\alpha\frac{d\theta}{\sqrt{1-m\sin^2\theta  }}\,,
\end{equation}
and where $r_{ij}=r_i-r_j$. Analogously the falling angle from the internal turning point to the event horizon $r_H=Q$ is
\begin{equation}\label{fal}
    \Delta\Phi(r_2,Q)=-\frac{2b}{\sqrt{r_{13}r_{24}}}F\Bigg[\arcsin\sqrt{\frac{r_{13}(r_2-Q)}{r_{23}(r_1-Q)}},\frac{r_{23}r_{14}}{r_{13}r_{24}}\Bigg]\,.
\end{equation}
In PM-expanded form these two angles are 
\begin{widetext}
\begin{align}
\chi(\infty,r_1)&=\pi\Bigg[\frac{Q^2 \left(6 E^2-\mu
   ^2\right)}{2 b^2 \left(E^2-\mu
   ^2\right)}+\frac{3 Q^4 \left(70 E^4-30 E^2
   \mu ^2+\mu ^4\right)}{8 b^4
   \left(E^2-\mu ^2\right)^2}+\frac{5 Q^6 \left(924 E^6-630 E^4
   \mu ^2+84 E^2 \mu ^4-\mu
   ^6\right)}{16 b^6 \left(E^2-\mu
   ^2\right)^3}\Bigg]\nonumber\\
   &+\frac{Q \left(4 E^2-2 \mu
   ^2\right)}{b \left(E^2-\mu
   ^2\right)}+\frac{2 Q^3 \left(40 E^6-72 E^4
   \mu ^2+33 E^2 \mu ^4-2 \mu
   ^6\right)}{3 b^3 \left(E^2-\mu
   ^2\right)^3}+\frac{1}{15 b^5
   \left(E^2-\mu ^2\right)^5}\Big(2 Q^5 \left(2016 E^{10}\right.\nonumber\\
   &\left.-6160E^8 \mu ^2 +6670 E^6 \mu ^4-2955
   E^4 \mu ^6+440 E^2 \mu ^8-8 \mu
   ^{10}\right)\Big)+O\left(\frac{1}{b^7}\right)\,,\nonumber\\
\chi(Q,r_2)&= \pi\Big[\frac{Q^2 \left(6 E^2-\mu
   ^2\right)}{2 b^2 \left(E^2-\mu
   ^2\right)}+\frac{3 Q^4 \left(70 E^4-30 E^2
   \mu ^2+\mu ^4\right)}{8 b^4
   \left(E^2-\mu ^2\right)^2}+\frac{5 Q^6 \left(924 E^6-630 E^4
   \mu ^2+84 E^2 \mu ^4-\mu
   ^6\right)}{16 b^6 \left(E^2-\mu
   ^2\right)^3}\Big] \nonumber\\
   &+\frac{4 E Q}{b \sqrt{E^2-\mu ^2}}+\frac{8 Q^3 \left(10 E^3-3 E \mu
   ^2\right)}{3 b^3 \left(E^2-\mu
   ^2\right)^{3/2}}+\frac{4 Q^5 \left(1008 E^5-560 E^3
   \mu ^2+45 E \mu ^4\right)}{15
   b^5 \left(E^2-\mu
   ^2\right)^{5/2}}+O\left(\frac{1}{b^7}\right)\,.
\end{align}
\end{widetext}
Comparing the arguments of the arcsin in \eqref{scatt} and \eqref{fal}, it is easy to show that
\begin{equation}
    \frac{r_{24}}{r_{14}}=\frac{r_{13}(r_2-Q)}{r_{23}(r_1-Q)}
\end{equation}
if $\mu=0$. This result follows from the Couch-Torrence (CT) symmetry, which implies that the scattering and infall angles of massless scalar probes coincide. Originally identified as a Weyl symmetry of the metric \cite{Couch:1984tr}, CT symmetry maps the horizon to infinity while leaving the photon sphere invariant, i.e., the locus of unstable circular null geodesics that massless probes approach asymptotically. For further details on this symmetry, see Refs. \cite{Cvetic:2020kwf,Cvetic:2021lss,Bianchi:2021yqs,Bianchi:2022wku,Akhond:2026suo}. The relevance of CT symmetry in the present context becomes manifest in the structure of the In and Up solutions, as discussed at the end of Subsec.~\ref{Upsec}.

\subsection{Circular geodesics}
Circular geodesics are determined by the conditions that the radial velocity and radial acceleration vanish simultaneously
\begin{equation}\label{circorb}
    Q_r(r_c,b_c)= \partial_r Q_r(r_c,b_c)=0\,,
\end{equation}
where the radial potential appears in \eqref{geoeq}. Solving the preceding condition yields the critical radius and the corresponding impact parameter
\begin{align}
    r_c&{=}\frac{Q\Big(2\Delta E^{2/3}{+}4E^2{-}6\mu^2{+}\frac{2E^{4/3}(4E^2{-}3\mu^2)}{\Delta}\Big)}{6(E^2{-}\mu^2)}\,,\nonumber\\
    b_c&{=}\frac{\mu  Q \sqrt{r_c} \sqrt{r_c
   \left(2 E^2{+}\mu ^2\right){-}\mu
   ^2 Q}}{\sqrt{E^2{-}\mu ^2}
   \sqrt{r_c{-}Q} \sqrt{2 \mu ^2 r_c{-}2 E^2
   r_c{+}4 E^2 Q{-}3 \mu
   ^2 Q}}\,,\nonumber\\
   \Delta^3&=8E^4+\frac{3}{2}E^2(\delta-15\mu)\mu-\frac{3}{2}(\delta-9\mu)\mu^3\nonumber\,,\\
   \delta^2&=81\mu^2-96E^2\,.
\end{align}
In the massless probe case the previous quantities reduce to
\begin{equation}
    r_c=2Q,\quad b_c=4Q\,.
\end{equation}
The dependence on the speed of light is restored knowing that
$Q/r \sim v^{2}$ and so $Q \to Q \eta^{2}$ where $\eta=1/c$. Similarly $\omega r \sim v $ and consequently $\omega \to \omega \eta$. In this notation the PN expansion consists in expanding physical quantities in series of $\eta$.
\section{Wave equation}
\label{waveEQ}
The equation for scalar waves
\begin{equation}
    \Box \Phi=0,\quad \Box=\frac{1}{\sqrt{-g}}\partial_\mu(\sqrt{-g}g^{\mu\nu}\partial_\nu)\,,
\end{equation}
can be separated through the ansatz 
\beq\label{anssf}
\Phi(t,r,\theta,\phi)=\int\frac{d\omega}{2\pi}e^{-i\omega t}R_{\ell m\omega}(r)Y_{\ell m}(\theta,\phi)\,,
\eeq
where $Y_{\ell m}(\theta,\phi)$ are the spherical harmonics on the sphere $S_2$ defined in terms of the Legendre polynomials $P_{\ell m}(\theta)$ as
\beq
Y_{\ell m}(\theta,\phi)=P_{\ell m}(\theta)e^{i m \phi }\,.
\eeq
After the separation we have 
\begin{align}\label{eqR}
&\frac{1}{\sin\theta}\frac{d}{d\theta}\left( \sin\theta \frac{d P_{\ell m}(\theta)}{d\theta}\right){-}\frac{m_\phi^2}{\sin^2\theta}P_{\ell m}(\theta){=}{-}\ell(\ell{+}1)P_{\ell m}(\theta)\nonumber\\
&\Bigg[(r{-}Q)\frac{d^2}{dr^2}{+}2\frac{d }{dr}{+}\frac{r^4\omega^2{-}\ell(\ell{+}1)(r{-}Q)^2}{(r{-}Q)^3}\Bigg]R_{\ell m\omega}(r){=}0\,. 
\end{align}
The computational complexity is entirely encoded in the radial equation, which takes the form of a DCHE, for which no explicit closed-form solution is known. In the following, we present two equivalent methods for constructing an expansion of the solutions to the corresponding homogeneous equation, which will turn out to yield the PN expansion. 
The first approach relies on the method of Floquet theory and kink transformations. The second is inspired by the MST method and suitably adapted to the DCHE to construct the solution in a complementary, yet equivalent, framework.

\subsection{From the DCHE to the radial equation}
As well known, the radial equation in \eqref{eqR} is a particular case of general DCHE
\begin{equation}
    \psi''(y)=V(y)\psi(y)\,,
\end{equation}
with potential
\begin{equation}
    V(y)= e^{2\theta}\left(e^{2y}+e^{-2y}\right)+e^\theta(2 q_1e^y+2 q_2e^{-y})+P^2\,,
    \label{VDCHE}
\end{equation}
via the map
\begin{align}
    & y= \log\left(\frac{r}{Q}-1 \right)\,, \notag \\
    &\psi(y)=\sqrt{\frac{r}{Q}-1} \ R(r)\,, \notag \\
    &q_1= - 2 i Q \omega, \notag \\
    &q_2= - 2 i Q \omega,  \notag\\
    &e^\theta=- i Q \omega\,, \notag \\
    &P= \frac{1}{2}\sqrt{(1+2\ell)^2-24 Q^2 \omega^2} \,.
    \label{gravitymap}
\end{align}
As anticipated, the general DCHE \eqref{VDCHE} admits on one side an interpretation as quantum integrable model, and on the other as $\mathcal{N}_{f}=2$ flavours superymmetric $SU(2)$ gauge theory, since its the null condition of the square of Seiberg-Witten differential deformed in the Nekrasov-Shatashvili (NS) regime \cite{Nekrasov:2009rc}, with flavour masses $m_{j}= \hbar q_{j}$, modulus $u= P^{2}/\hbar^{2}$, and instanton coupling $\Lambda_{2}= 4 \hbar e^{\theta}$\cite{Fioravanti:2021dce}.
In this work we are particularly interested in the incoming (In) and the outgoing (Up) solutions of the DCHE. The In solution is uniquely defined by the following subdominant behaviour for $y \to -\infty$:
\beq
\psi_{\rm in}(y) \simeq  \exp\left[(-\frac{1}{2} -q_{2})(\theta-y+\ln2)-e^{\theta-y}\right]\,,
\label{asyIN}
\eeq
while the Up solution is uniquely defined by the following subdominant behaviour for $y \to +\infty$:
\beq
\psi _{\rm up}(y)\simeq \exp \left [(-\frac {1}{2}- q_1)(\theta + y +\ln 2)-e^{\theta + y} \right]\,.
\label{asyUP}
\eeq
From \eqref{gravitymap} the limit $y \to -\infty$ would correspond to $r \to Q$ and so 
\begin{equation}
  R_{\rm in}(r)=\mathcal{N}_{\rm in} \frac{1}{\sqrt{\frac{r}{Q}-1}} \psi_{\rm in}(y)  \,,
  \label{Rpsi}
\end{equation}
with $ \mathcal{N}_{\rm in}$ a suitable normalization constant 
satisfies the incoming boundary conditions at the horizon, while 
 $y\to +\infty$ would correspond to $r \to \infty$ and so 
 \begin{equation}
  R_{\rm up}(r)= \mathcal{N}_{\rm up}\frac{1}{\sqrt{\frac{r}{Q}-1}} \psi_{\rm up}(y)  \,,
  \label{Rpsiupbis}
\end{equation}
with again $ \mathcal{N}_{\rm up}$ a normalization constant satisfies outgoing boundary conditions at infinity. \\
The potential (\ref {VDCHE}) allows various symmetries. Two of them are important for our aims. The first one, $\hat \Omega _1$, sends $P\rightarrow -P$ and does not affect $\psi_{\rm in/up}(y)$: $\psi_{\rm in/  up}(y,-P) = \psi_{\rm in/up}(y,P)$. The second one, $\hat{\Omega} _2$, operates as
$y\rightarrow -y, q_1 \leftrightarrow q_2$ and then exchanges $\psi_{\rm in}(y)$ with $\psi_{\rm up}(y)$:
$\psi_{\rm in}(-y,q_2,q_1) = \psi_{\rm up}(y,q_1,q_2)$.  \\
Another important set of solutions is given by the Floquet solutions $\psi_\pm(y)$ defined by the quasi-periodicity property
\begin{equation}
    \psi_\pm(y+2 \pi i) = e^{\pm2 \pi i k }\psi_\pm(y)\,,
\end{equation}
with $k$ called Floquet index (or exponent).\\
In general, we can consider the Eikonal form of the wave function 
\begin{equation}
    \psi(y) = \exp\left( \int^y \Pi(y') dy' \right) \,,
    \label{psiEikonal}
\end{equation}
where $\Pi(y)$ is the quantum momentum which solves the Riccati equation
\begin{equation}
    \Pi(y)^2+\frac{d}{dy} \Pi(y) =V(y)\,.
    \label{riccati}
\end{equation}
We can then call $\Pi_{ \rm in}(y)$,  $\Pi_{up}(y)$ and  $\Pi_{\pm}(y)$
the quantum momentum associated to $\psi_{\rm in}(y)$,  $\psi_{\rm up}(y)$ and  $\psi_{\pm}(y)$ respectively.
\subsection{Large $\theta\to-\infty$: Floquet index}
A central role in our construction is played by the Floquet index $k$, computed as
\beq
k= \frac{1}{2 \pi i } \int_y^{y+2 \pi i } dy' \Pi_+(y')\,.
\label{Floquet}
\eeq
In gravity setting we would be interested in constructing some observables, like the energy loss, in the PN regime defined by 
\beq
Q \mapsto Q \eta^{2}, \qquad \omega \mapsto \omega \eta, \qquad \eta \to 0\,.
\label{PNdef}
\eeq
Due to the map \eqref{gravitymap} this regime can be studied in the DCHE through the $\theta \to -\infty$ limit. Following the aforementioned  \cite{Fioravanti:2025bts}, we can write the quantum momentum as a convergent series in $e^{\theta}$
\begin{equation}
    \Pi_+(y)=\sum_{n=0}^{+\infty}\Pi_+^{(n)}(y) e^{n\theta}\,.
    \label{Pin1}
\end{equation}
Plugging this ansatz into the Riccati equation \eqref{riccati} we have the set of equations
\begin{equation}
    \sum_{m=0}^n\Pi _+^{(m)}(y)\Pi _+^{(n-m)}(y)+\frac{d}{dy}\Pi_+^{(n)}(y)=V^{(n)}(y)\,,
\end{equation}
where $V^{(0)}(y)=P^2$, $V^{(1)}(y)=2q_1e^y+2q_2e^{-y}$, $V^{(2)}(y)=2\cosh 2 y$, $V^{(n)}(y)=0$ for $n\geq 3$. \\
For generic $P$, the solution can be written as
\begin{equation}
    \Pi_+^{(n)}(y) = \sum_{m=-n}^{n} r_m^{(n)} e^{my}\,,
\end{equation}
and the constants satisfy the recursion relations \cite{Fioravanti:2025bts}
\begin{widetext}
\begin{align}
&r_0^{(0)} = P\,, \\[0.6em]
&p r_p^{(1)} + \sum_{m=0}^{1} \sum_{h=\max\{-m,\, p+m-1\}}^{\min\{m,\, p-m+1\}} r_h^{(m)} r_{p-h}^{(1-m)} = 2q_1 \delta_{p,1} + 2q_2 \delta_{p,-1}\,, \quad -1 \le p \le 1\,, \\[0.6em]
&p r_p^{(2)} + \sum_{m=0}^{2} \sum_{h=\max\{-m,\, p+m-2\}}^{\min\{m,\, p-m+2\}} r_h^{(m)} r_{p-h}^{(2-m)} = \delta_{p,-2} + \delta_{p,2}\,, \quad -2 \le p \le 2\,, \\[0.6em]
&p r_p^{(n)} + \sum_{m=0}^{n} \sum_{h=\max\{-m,\, p+m-n\}}^{\min\{m,\, p-m+n\}} r_h^{(m)} r_{p-h}^{(n-m)} = 0\,, \quad -n \le p \le n\,, \quad n \ge 3\,.
\end{align}
\end{widetext}
The solutions at first orders in $e^{\theta}$ are
\begin{align}
    &\Pi_+^{(0)}(y)=P\,,
    \\
    &\Pi_+^{(1)}(y)=\frac{2q_2e^{-y}}{2P-1}+\frac{2q_1e^{y}}{2P+1}\,,
    \\
    &\Pi_+^{(2)}(y)=-\frac{(2P-1-2q_2)(2P-1+2q_2)e^{-2y}}{(2P-1)^2(2-2P)}\nonumber \\
    &+\frac{4q_1q_2}{P(1-4P^2)}+\frac{(2P+1-2q_1)(2P+1+2q_1)e^{2y}}{(2P+1)^2(2+2P)}\,.
\end{align}
In this way the Floquet index \eqref{Floquet} is given by
\begin{widetext}
\begin{align}
    k&=P+ \frac{4 q_1 q_2}{P(1 - 4 P^2)}  e^{2\theta}  +\frac{-P^2 (4 P^2 - 1)^3 + 12 P^2 (4 P^2 - 1)^2 (q_1^2 + q_2^2) - 16 (2 - 35 P^2 + 60 P^4) q_1^2 q_2^2}{4 P^3 (P^2 - 1) (4 P^2 - 1)^3}     e^{4\theta} +O(e^{6 \theta}) \,,
    \label{FlouqetNf2}
\end{align}
\end{widetext}
and it receives contributions only from constant terms of even $n$. 
This relation can also be eventually inverted to obtain
\begin{equation}
    P^2=\sum_{n=0}^{+\infty}p_2^{(n)}(k)e^{2n\theta}=k^2+\frac{8 q_{1} q_{2}}{4k^{2}-1}e^{2 \theta}+O(e^{4\theta}) \,.
    \label{P2k}
\end{equation}
Now, if we try to express the series (\ref {FlouqetNf2}) in gravity variables, through the map (\ref {gravitymap}), we find the following phenomenon. When $\ell \geq 1$ the computation of $\Pi_+^{(n)}(y)$ up to $n=5$ is sufficient to determine the Floquet exponent in gravitational variables as a series in $(Q\omega)^2$, up to the order $(Q\omega )^8$:
\begin{widetext}
\begin{align}
    k^{(\ell=1)}&=\frac{3}{2}-2Q^2\omega^2-\frac{14Q^4\omega^4}{5}-\frac{2144Q^6\omega^6}{225}-\frac{67706Q^8\omega^8}{1575}+O\Big(Q^{10}\omega^{10}\Big)\,,\nonumber\\
    k^{(\ell=2)}&=\frac{5}{2}-\frac{6 Q^2\omega ^2}{5}-\frac{502 Q^4\omega ^4}{875}-\frac{294004Q^6 \omega^6}{459375}-\frac{15685184 Q^8 \omega^8}{16078125}+O\Big(Q^{10}\omega^{10}\Big)\,,\nonumber\\
    k^{(\ell=3)}&=\frac{7}{2}-\frac{6Q^2\omega^2}{7}-\frac{3188Q^4\omega^4}{15435}-\frac{1296541Q^6\omega^6}{11344725}-\frac{13985769332Q^8\omega^8}{165099782925}+O\Big(Q^{10}\omega^{10}\Big)\,,\nonumber\\
    k^{(\ell=4)}&=\frac{9}{2}-\frac{2Q^2\omega^2}{3}-\frac{3016Q^4\omega^4}{31185}-\frac{10381141Q^6\omega^6}{324168075}-\frac{123632605378Q^8\omega^8}{8761290563025}+O\left(Q^{10}\omega^{10}\right)\,.
    \label{RAM}
\end{align}
\end{widetext}
When $\ell =0$, however, the use of \eqref{FlouqetNf2} to find the Floquet index in gravity variables is more involved, since our computations suggest that an infinite number of terms in the expansion \eqref{FlouqetNf2} contribute at every given order $(\omega Q)^{2n}$. Then, an easier way to calculate $k^{(\ell=0)}$ is trying to solve the Riccati equation \eqref{riccati} directly in gravity variables (with $\ell=0$):
\begin{equation}
    (\Pi^{(\ell=0)})'(y)+(\Pi^{(\ell=0)}(y))^2
    =\frac{1}{4}-(\omega Q)^2 W(y),
    \label{eq:riccati}
\end{equation}
where
\begin{equation}
    W(y)=e^{2y}+4e^y+6+4e^{-y}+e^{-2y}
    =\frac{(e^y+1)^4}{e^{2y}}.
    \label{eq:source}
\end{equation}
The Floquet index is given by
\begin{align}
k^{(\ell=0)}&= \frac{1}{2 \pi i } \int_y^{y+2 \pi i } ds \Pi^{(\ell=0)}(s) 
\label{Floquet1}\,.
\end{align}
We seek a \(2\pi i\)-periodic solution of \eqref{eq:riccati} in the series form \footnote{These $\Pi_{2n}^{(\ell=0)}(y)$ are different from the $\Pi_+^{(n)}(y) $ of \eqref{Pin1} expressed in gravity variables.}
\begin{equation}
    \Pi^{(\ell=0)}(y)=\Pi_0^{(\ell=0)}(y)+\sum_{n=1}^{\infty}\Pi_{2n}^{(\ell=0)}(y) (\omega Q)^{2n}
    \label{PiwQ}
\end{equation}
and then require each function $\Pi_{2n}^{(\ell=0)}(y)$ to be \(2\pi i\)-periodic. In summary, we found (see Appendix \ref{FloquetL0})
\begin{align}
&\Pi_0^{(\ell=0)}(y)=\frac{1}{2} \left( \frac{e^{y}-e^{y_{0}}}{e^{y}+e^{y_{0}}} \right),
\end{align}
with integration constants
\begin{equation}
   e^{y_0}=\frac{-3\pm\sqrt{5}}{2}\,,
   \label{y0}
\end{equation}
and
\beq
\Pi^{(\ell=0)}_{2n}(y) = \frac{\sum\limits_{m=0}^{6n} g_{n,m}\, e^{m y}}{e^{2n y}(e^y + e^{y_{0}})^{2n}} \qquad \text{for } n = 1, 2, 3,\dots
\label{PiL0}
\eeq
where the $g_{n,m}$ are computed order by order.\\
In \eqref{y0} the upper sign, through \eqref{Floquet1},  gives the positive Floquet $k$, while the lower sign would give $-k$.\\
Finally \eqref{Floquet1} gives
\beq
k^{(\ell=0)}= \frac{1}{2} + \sum_{n=1}^{\infty}\Delta k_{2n} (\omega Q)^{2n}
\eeq
where
\beq
\Delta k_{2n} \equiv \frac{1}{2\pi i} \int_0^{2\pi i} \Pi^{(\ell=0)}_{2n}(y)\, dy,  \qquad \text{for } n \in \mathbb{N}-\{0\}
\eeq
which can be calculated using the residue theorem.
We calculated up to $ \Pi^{(\ell=0)}_{8}(y)$ to obtain 
\bea
&k^{(\ell=0)}=\frac{1}{2}-2 \sqrt{5} Q^2 \omega ^2-\frac{42 Q^4 \omega^4}{\sqrt{5}}-\frac{17536 Q^6 \omega ^6}{45
   \sqrt{5}} \nonumber \\
  & -\frac{943054Q^8\omega^8}{225\sqrt{5}} 
   +O\Big(Q^{10}\omega^{10}\Big) \, .
\eea

\subsection{The kink transformation}
Looking at the potential \eqref{VDCHE} it is evident the incompatibility between the limits $\theta \to -\infty$ and $y \to \pm \infty$. 
In order to overcome this issue we will have to do the so-called \textit{kink transformation} \cite{Fioravanti:2025bts}. For $y<0$ we should do the shift $y\to y+\theta$ and define the \textit{left kink} functions $\psi_{<}(y)=\psi(y+\theta)$, $\Pi_{<}(y)=\Pi(y+\theta)$ and $V_{<}(y)=V(y+\theta)$, while for $y>0$ we should do the shift $y\to y-\theta$ and define the \textit{right kink} functions $\psi_{>}(y)=\psi(y-\theta)$, $\Pi_{>}(y)=\Pi(y-\theta)$ and $V_{>}(y)=V(y-\theta)$. 
\\
Consider the Riccati equation for  $\Pi_{ \rm in,<}(y)$ in the left kink variable
\begin{equation}\label{ricc<0}
    \Pi_{ \rm in,<}(y)^2+\frac{d}{dy}\Pi_{ \rm in,<}(y)=V_<(y)  \,,
\end{equation}
where $V_<(y) =\sum_{n=0}^{+\infty}  V^{(n)}_<(y) e^{2 n \theta}$ and we have used the expansion of $P^{2}$ in terms of $k$ \eqref{P2k}. Then, 
\begin{align}
    &V_<^{(0)}(y)=e^{-2y}+2q_2e^{-y}+k^2 \,,\\
    &V^{(1)}_<(y)=2q_1e^{y}+p_2^{(1)}(k) \,, \\
    &V^{(2)}_<(y)=e^{2y}+p_2^{(2)}(k) \,,\\
    &V^{(n)}_<(y)=p_2^{(n)}(k) \quad { \rm for} \quad n\geq 3 \,.
\end{align}
We can write the solution in series as
\begin{equation}
 \Pi_{ \rm in,<}(y)=\sum_{n=0}^{+\infty}\Pi_{ \rm in,<}^{(n)}(y)  e^{2n\theta} \,.
 \label{expPi}
 \end{equation}
 Putting then \eqref{expPi} in \eqref{ricc<0}
 at zero-th order in $e^{2\theta}$ we have
\begin{equation}
 \Pi_{ \rm in,<}^{(0)}(y)^2+\frac{d}{dy}\Pi_{ \rm in,<}^{(0)}(y) = V_<^{(0)}(y) \,.
 \label{Ricc0}
\end{equation}
The asymptotic behaviour of \eqref{asyIN} is implemented by choosing
\begin{equation}
   \Pi_{ \rm in ,<}^{(0)}(y)=\frac{d}{dy}\ln \left(e^{-e^{-y}+ky}U\left(\frac{1}{2}-k+q_2,1-2k,2e^{-y}\right)\right) \,,
   \label{Pi0IN}
\end{equation}
where $U$ is the Tricomi confluent hypergeometric function.
Instead, at orders $e^{2n\theta}$ we have the equations
\bea
&2\Pi_{ \rm in,<}^{(0)}(y) \Pi_{ \rm in,<}^{(n)}(y) + \frac{d}{dy} \Pi_{ \rm in,<}^{(n)}(y) \nonumber \\
&+ \sum\limits_{m=1}^{n-1} \Pi_{ \rm in,<}^{(m)}(y) \Pi_{ \rm in,<}^{(n-m)}(y) =V^{(n)}_{<}(y)\,,
\eea
where (see Appendix \ref{Ricc})  the solutions $\Pi_{ \rm in, <}^{(n)}(y)$ are given by 
\begin{widetext}
\begin{equation}
   \Pi _{in, <}^{(n)}(y) = \frac{d}{dy}\left[P^{(n)}_0(y;q_1,q_2,k)+  \sum\limits_{j=1}^nP_{j}^{(n)}(y;q_1,q_2,k)  \frac{d^{j-1}}{dy^{j-1}}\Pi_{ \rm in ,<}^{(0)}(y) \right ]\,,
   \label{Pin}
\end{equation} 
\end{widetext}
with 
\begin{align}
&P^{(n)}_0(y;q_1,q_2,k) = \sum_{m=1}^{n}l_{n0m}(q_1,q_2,k)e^{my}\,,  \label{Pn0}\\ 
    &P^{(n)}_j(y;q_1,q_2,k) =\sum_{m=j}^{n}l_{njm}(q_1,q_2,k)e^{my} \label{Pnj}\,.
\end{align}
As a consequence of the invariance of $\Pi_{ \rm in, <}(y)$ under symmetry $\hat \Omega _1$, 
one has that $\Pi_{ \rm in, <}(y,q_1,q_2,k)=\Pi_{ \rm in, <}(y,q_1,q_2,-k)$, since from \eqref{FlouqetNf2} it follows  that $P \to -P$ is equivalent to $k \to -k$.
The expression for the Up solution follows from that of the In solution by using the symmetry $\hat \Omega _2$, which implies 
\beq
 \Pi_{\rm up,>}(y,q_{1},q_{2},k)= - \Pi_{ \rm in,<}(-y,q_{2},q_{1},  k)\,.
 \label{Piinuprelation}
\eeq
\subsection{The DCHE wavefunction}
\label{DCHEwave}
For the In solution we have
\begin{equation}
    \psi_{\rm in}(y) = \psi_{\rm in,<}(y-\theta)\,,
    \label{psipsi<}
\end{equation}
where 
\begin{align}
\psi_{\rm in, <}(y)& =\mathcal{N} \exp\left( \int^y \Pi_{ \rm in,<}(y') dy' \right) = \nonumber \\
&= \mathcal{N} \exp\left(\sum_{n=0}^{\infty} e^{2 n \theta} \int^y \Pi^{(n)}_{in,<}(y') dy' \right)\,,
\end{align}
with $\mathcal{N}$ a constant, whose expression, chosen to fit the asymptotics (\ref {asyIN}), will be given below. Using  \eqref{Pin} and expanding for $e^{\theta} \to 0$ we obtain \cite{Fioravanti:2025bts}
\begin{widetext}
\begin{align}
\psi_{\rm in,<}(y) ={}&\mathcal{N} \Bigg(\left[ 1 + e^{2\theta} P_0^{(1)}(y) + e^{4\theta} \left( P_0^{(2)}(y) + \frac{1}{2} \left( P_0^{(1)}(y) \right)^2 \right) + O(e^{6\theta}) \right] \psi_{\rm in, <}^{(0)}(y) + {} \notag \\
&{}+ \left[ e^{2\theta} P_1^{(1)}(y) + e^{4\theta} \left( P_1^{(2)}(y) + P_0^{(1)}(y) P_1^{(1)}(y) \right) + O(e^{6\theta}) \right] \frac{d\psi_{\rm in,<}^{(0)}(y)}{dy} +\nonumber  \\
&{}+ \left[ e^{4\theta} \left( P_2^{(2)}(y) \frac{d}{dy}\Pi_{ \rm in,<}^{(0)}(y) + \frac{1}{2} \left( P_1^{(1)}(y) \right)^2 \left( \Pi_{ \rm in,<}^{(0)}(y) \right)^2 \right) + O(e^{6\theta}) \right] \psi_{\rm in,<}^{(0)}(y) \Bigg), 
\end{align}
\end{widetext}
where 
\bea
& \psi^{(0)}_{\rm in,<}(y)=\exp\left(\int^{y}\Pi^{(0)}_{\rm in,<}(y') \right)= 
\nonumber \\
&=e^{-e^{-y}}e^{ky} U\left(\frac{1}{2}-k+q_2,1-2k,2e^{-y}\right)\,.
\label{PsiIn}
\eea
Using \eqref{Ricc0} and the relation \eqref{P22} we can see that the coefficient of $\left(\Pi_{ \rm in,<}^{(0)}(y)\right)^2$ gets cancelled and we can group $\psi_{\rm in,<}^{(0)}(y) $ and its derivative as
\begin{widetext}
\begin{align}
\psi_{\rm in,<}(y) ={}&\mathcal{N} \Bigg( \left[ 1 + e^{2\theta} P_0^{(1)}(y) + e^{4\theta} \left( P_0^{(2)}(y) + \frac{1}{2} \left( P_0^{(1)}(y) \right)^2+P_2^{(2)}(y)V_{<}^{(0)}(y)  \right) + O(e^{6\theta}) \right] \psi_{\rm in,<}^{(0)}(y) + {} \notag \\
&{}+ \left[ e^{2\theta} P_1^{(1)}(y) + e^{4\theta} \left( P_1^{(2)}(y) + P_0^{(1)}(y) P_1^{(1)}(y) \right) + O(e^{6\theta}) \right] \frac{d\psi_{\rm in,<}^{(0)}(y)}{dy}  \Bigg) 
\label{Psiless}\,.
\end{align}
\end{widetext}
Analogous cancellations for powers $\left(\Pi_{ \rm in,<}^{(0)}(y)\right)^n$ happen at every order in $e^{2n \theta}$ (we checked these cancellations up to $e^{14 \theta}$), thanks to some identities among the polynomials $P^{(n)}_{m}(y)$ (see Appendix \ref{Ricc}), because of their apparent expressions in terms of Bell polynomials. This leads us to conjecture \footnote{Analogous method is applied in the right kinked case by replacing the left kinked functions with the right kinked ones.}  the exact form  
\begin{equation}
    \psi_{\rm in,<}(y) = \mathcal{N} \left( \mathcal{S}(y) \psi^{(0)}_{\rm in,<}(y) +   \mathcal{\hat{S}}(y) \frac{d\psi_{\rm in,<}^{(0)}(y)}{dy} \right)  \,,
    \label{psiS1}
\end{equation}
where
\begin{align}
& \mathcal{S}(y) =1+ \sum_{n=1}^{+\infty} \mathcal{S}_{n}(y) e^{2n\theta } \,,
\label{s0}\\
&\mathcal{\hat{S}}(y) =\sum_{n=1}^{+\infty} \mathcal{\hat{S}}_{n}(y) e^{2n \theta} \,.
\label{s1}
\end{align}
An analogous formula to \eqref{psiS1} will be proved for the Confluen Heun Equation (CHE) in an upcoming publication \cite{FRO} and generalized to all the other Heun Equations in \cite{FRO2}.\\
From \eqref{Pn0}-\eqref{Pnj} and \eqref{Psiless} one can see that also $S_{n}(y)$ and $\hat{S}_{n}(y)$ are polynomials in $e^{y}$ of degree $n$ (see Appendix \ref{AppS})
\bea
&S_{n}(y) &= \sum_{m=0}^{n}s_{n,m}e^{my} \,, \label{Sexp}\\
&\hat{S}_{n}(y) &=  \sum_{m=1}^{n}\hat{s}_{n,m}e^{my}\,. \label{Shexp} 
\eea
Then, in order to satisfy the asymptotic expansion (\ref {asyIN}), the normalization constant has to be equal to
\beq
\mathcal{N}=2^{-k} \left [1+\sum _{n=1}^{+\infty}\left (s_{n,0}+\hat s_{n,1} \right ) e^{2n\theta} \right]^{-1}\,.
\label{calN}
\eeq
Then, the final expression for $\psi_{\rm in,<}(y)$ is
\begin{equation}
    \psi_{\rm in,<}(y) = \frac {\left( \mathcal{S}(y) \psi^{(0)}_{\rm in,<}(y) +   \mathcal{\hat{S}}(y) \frac{d\psi_{\rm in,<}^{(0)}(y)}{dy} \right)}  {   2^k \left [1+\sum\limits _{n=1}^{+\infty}\left (s_{n,0}+\hat s_{n,1} \right ) e^{2n\theta} \right] }\, .
    \label{psiS1-final}
\end{equation}
Obviously, inserting  \eqref{psipsi<}  into  \eqref{Rpsi} gives the final expression 
\begin{equation}
  R_{\rm in}(r)= \mathcal{N}_{\rm in}\frac{1}{\sqrt{\frac{r}{Q}-1}} \psi_{\rm in,<}(y-\theta)  \,,
  \label{Rpsibis}
\end{equation}
with $\psi_{\rm in,<}(y)$ expressed as \eqref{psiS1}. Now, with the integrability-gravity map \eqref{gravitymap}  we can, for instance, write down its PN expansions for different $\ell$ as reported in Appendix \ref{appaA}. With the same passages we can write the Up solution as
\begin{equation}
  R_{\rm up}(r)=  \mathcal{N}_{\rm up}\frac{1}{\sqrt{\frac{r}{Q}-1}} \psi_{\rm up,>}(y+\theta)  \,.
  \label{Rpsi_up}
\end{equation}
In order to reconstruct $\psi_{\rm up,>}(y) $ we remark that the use of \eqref{Piinuprelation} together with asymptotic behaviours (\ref {asyIN}, \ref {asyUP}) fixes the following relation between the kinked wave functions
\bea
\psi_{\rm up,>}(y,q_1,q_{2},k)  =  \psi_{\rm in,<}(-y,q_2,q_{1}, k)\, .
\label{psiCT0}
\eea
We can use then the $\hat{\Omega}_{1}$ symmetry to write
\bea
\psi_{\rm up,>}(y,q_1,q_{2},k)  =  \psi_{\rm in,<}(-y,q_2,q_{1}, -k)\, ,
\label{psiCT}
\eea
which gives rise to the CT symmetry once the parameters are restricted to gravity: this is shown below at the end of Subsec. \ref {Upsec}. The evident advantage of our new solution \eqref{psiS1-final} or \eqref{Rpsibis}  is to have only one hypergeometric function, and, in principle, for future applications, one can try to integrate directly on it: in fact, this is a clear advantage w.r.t. the MST method ({\it cf.} next section). For the time being, though, we follow the traditional method and expand in PN approximation the unique hypergeometric function and the $\mathcal{S}$, $\mathcal{\hat{S}}$ in (\ref{psiS1-final}) into the usual functions, reported in Appendix \ref{appaA}. 
Eventually these building blocks can all be integrated on to obtain our final goals. Nevertheless, to check our results \eqref{Rpsibis} and \eqref{Rpsi_up},  we are now going to match them in details with the MST ones. This is indeed possible as already anticipated by \cite{Fioravanti:2025bts} for other Heun-like equations (Modified Mathieu Equation and Confluent Heun Equation).
  
\subsection{Kink and MST methods in comparison}\label{FloqMST}

The function \eqref{PsiIn}, due to some shift properties of the Tricomi $U$ function \cite{Fioravanti:2025bts}, satisfies the relations
\begin{widetext}
\beq
 {e^{y}} \psi _{\rm in,<}^{(0)}(y;k)=- \frac {2(\frac {1}{2}-k+q_2)}{k(1-2k)}\psi _{\rm in,<}^{(0)}(y;k-1) + \frac {q_2}{\frac {1}{4}-k^2}\psi _{\rm in,<}^{(0)}(y;k) + \frac {\frac {1}{2}+k+q_2}{2k(1+2k)}\psi _{\rm in,<}^{(0)}(y;k+1) \label {shift-1} \, , 
\eeq
\beq
e^{y}\frac {d}{dy} \psi _{\rm in,<}^{(0)}(y;k)= \frac {2(\frac {1}{2}-k+q_2)}{1-2k}\psi _{\rm in,<}^{(0)}(y;k-1) - \frac {q_2}{\frac {1}{2}-2k^2}\psi _{\rm in,<}^{(0)}(y;k) +\frac {\frac {1}{2}+k+q_2}{2(1+2k)}\psi _{\rm in,<}^{(0)}(y;k+1)  \label {shift-2} \, . 
\eeq
\end{widetext}
We can use these identities iteratively to write
\bea
&e^{my} \psi_{\rm in,<}^{(0)}(y; k) = \sum\limits_{q=-m}^{m} \alpha_{m, q}(k) \, \psi_{\rm in,<}^{(0)}(y; k + q) \,,\nonumber \\
&e^{my} \frac{d}{dy} \psi_{\rm in,<}^{(0)}(y; k) = \sum\limits_{q=-m}^{m} \beta_{m, q}(k) \, \psi_{\rm in,<}^{(0)}(y; k + q) \,.\nonumber \\
\eea
Applying them to \eqref{psiS1-final} we can write the $\psi_{\rm in,<}$ in a MST like form
\bea
\psi_{\rm in,<}(y; k) = \frac { \sum\limits_{n = -\infty}^{+\infty} C_n(k; e^{2\theta}) \, \psi_{\rm in,<}^{(0)}(y; k + n)} {   2^k \left [1+\sum _{n=1}^{+\infty}\left (s_{n,0}+\hat s_{n,1} \right ) e^{2n\theta} \right] }\,,
\label{psiLess2}
\eea
where the coefficients are given by
\begin{widetext}
\begin{align}
C_n(k;e^{2\theta})
={}&\delta_{n,0}+\sum_{j=1}^{\infty}e^{2 j \theta}
\left[
\sum_{m=0}^{j}s_{j,m}\alpha_{m,n}(k)
+
\sum_{m=1}^{j}\hat{s}_{j,m}\beta_{m,n}(k)
\right].
\end{align}
\end{widetext}
Finally the expression for the radial incoming solution (\ref {Rpsi}) is 
\beq
R_{ \rm in}(r)=\mathcal{N}_{\rm in}e^{-\frac {y}{2}}  \frac { \sum\limits_{n = -\infty}^{+\infty} C_n(k; e^{2\theta}) \, \psi_{\rm in,<}^{(0)}(y-\theta; k + n)} {   2^k \left [1+\sum\limits_{n=1}^{+\infty}\left (s_{n,0}+\hat s_{n,1} \right ) e^{2n\theta} \right] } \,.
\label {RpsiLess2}
\eeq
If we introduce the following coordinate in the radial wave equation \eqref{eqR}
\begin{equation}
    z=r-Q= e^{y}Q \,,
\end{equation}
we can write, using \eqref{gravitymap},
\begin{widetext}
\begin{equation}\label{eq:dizionario}
\psi^{(0)}_{\rm in,<}(y-\theta;k+n)=e^{\frac{iQ^2\omega}{z}}(-iQ^2\omega)^{-k-n}\;
z^{k+n}\,U\!\Big(\frac{1}{2}-k-n-2iQ\omega,1-2k-2n,-\tfrac{2iQ^2\omega}{z}\Big).
\end{equation}
\end{widetext}
All the first coefficients $C_n$ can be computed explicitly and the first ones are 
\begin{align}
&C_0(k; e^{2\theta})=1+\frac{16q_1q_2}{(1-4k^2)^2}e^{2\theta}+O(e^{4\theta}),\qquad \\
&C_1(k; e^{2\theta})=\frac{q_1(\frac12+k+q_2)}{k(1+2k)^2}e^{2\theta}+O(e^{4\theta}),\\
&C_{-1}(k; e^{2\theta})=-\frac{4q_1(\frac12-k+q_2)}{k(1-2k)^2}e^{2\theta}+O(e^{4\theta}).
\end{align}
 In the following we are going to verify analytically our new method vs. the traditional MST approach. 
\subsection{MST: In solution}
\label{MSTin}
In order to find the incoming radial wave function it is convenient to write it as 

\begin{equation}
    R_{\rm in}(z)=e^{i\omega z}e^{\frac{i\omega Q^2}{z}}\frac{1}{z}f(z)\,.
    \label{RinMST}
\end{equation}
So that, Eq. \eqref{eqR} becomes
\begin{align}\label{fofz}
    &0=z^2f''(z){-}2i\omega(Q^2{-}z^2)f'(z)\nonumber\\
    &{+}\Big[{-}\ell(\ell{+}1){+}8Q^2\omega^2{+}4Q\omega^2z{+}\frac{2(iQ^2\omega{+}2Q^3\omega^2)}{z}\Big]f(z)\,.
\end{align}
Let us split \eqref{fofz} as follows
\begin{align}\label{ineq}
&z^2f''(z){-}2iQ^2\omega f'(z){+}\Big[{-}\nu{-}\nu^2{+}\frac{2Q^2\omega(i{+}2Q\omega)}{z}\Big]f(z)\nonumber\\
&{=}\Big[\ell{+}\ell^2{-}\nu{-}\nu^2{-}8Q^2\omega^2{-}4Q\omega^2 z\Big]f(z){-}2i\omega z^2f'(z)
\end{align}
where we introduce the Renormalized Angular Momentum (RAM) $\nu$.

The RAM is introduced ad hoc and is to be understood as satisfying $\nu=\ell+\mathcal{O}(Q^{2}\omega^{2})$. Hence, $\nu-\ell=\mathcal{O}(Q^{2}\omega^{2})$ is a small quantity in the regime $Q\omega\ll1$. This choice ensures that the entire right-hand side of the equation is proportional to $Q\omega$, allowing it to be treated as a perturbatively small source.

The left-hand side of \eqref{ineq} can be solved in terms of confluent hypergeometric functions requiring ingoing boundary conditions at the horizon $z=0$
\begin{align}
    f_\nu(z)=z^{1{+}\nu}U\Big({-}\nu{-}2iQ\omega,{-}2\nu,{-}\frac{2iQ^2\omega}{z}\Big)\,,
    \label{fnu}
\end{align}
where $U$ again is the Tricomi confluent hypergeometric function. Our ansatz for ingoing solution of \eqref{ineq} is
\begin{align}
    f(z)=\sum_{n=-\infty}^\infty c_n f_{\nu+n}(z)\,.
    \label{fz}
\end{align}
The right-hand side can be treated using the following identities
\begin{align}
z^2f'_{n+\nu}(z)=A_1 f_{n+\nu+1}(z)+A_0 f_{n+\nu}(z)+A_{-1} f_{n+\nu-1}(z)\,,\nonumber\\
zf_{n+\nu}(z)=B_1 f_{n+\nu+1}(z)+B_0 f_{n+\nu}(z)+B_{-1} f_{n+\nu-1}(z)\,,
\end{align}
where
\begin{align}
A_1&=\frac{1+n+\nu-2iQ\omega}{2(1+2n+2\nu)}\,,\nonumber\\
A_0&=iQ^2\omega\,,\nonumber\\
A_{-1}&=-\frac{2Q^4\omega^2(n+\nu+2iQ\omega)}{1+2n+2\nu}\,,\nonumber\\
B_1&=\frac{1+n+\nu-2iQ\omega}{2(1+n+\nu)(1+2n+2\nu)}\,,\nonumber\\
B_0&=\frac{2Q^3\omega^2}{(n+\nu)(n+\nu+1)}\,,\nonumber\\
B_{-1}&=\frac{2Q^4\omega^2(n+\nu+2iQ\omega)}{(n+\nu)(1+2n+2\nu)}\,.
\end{align}
Using the previous identities Eq. \eqref{ineq} can be rewritten as a three-terms recursion of the form
\begin{align}
   &\frac{i\omega(1+n+\nu-2iQ\omega)^2}{(1+n+\nu)(1+2n+2\nu)}f_{n+\nu+1}(z)\nonumber\\
   &+\Big(n(n+1)-\ell(\ell+1)+\nu(\nu+1)\nonumber\\
   &+2n\nu+6Q^2\omega^2+\frac{8Q^4\omega^4}{(n+\nu)(1+n+\nu)}\Big)f_{n+\nu}(z)\nonumber\\
   &-\frac{4iQ^4\omega^3(n+\nu+2iQ\omega)^2}{(n+\nu)(1+2n+2\nu)}f_{n+\nu-1}(z)=0\,.
\end{align}
The coefficients of the previous recursion can be shifted to obtain
\begin{align}\label{3terms}
    &\alpha_n c_{n+1}+\beta_n c_n+\gamma_{n}c_{n-1}=0\,,\nonumber\\
    &\alpha_n=-\frac{4iQ^4\omega^3(1+n+\nu+2iQ\omega)^2}{(1+n+\nu)(3+2n+2\nu)}\,,\nonumber\\
   &\beta_n=-\ell (\ell+1)+\nu(\nu+1) +n(n+1)+2 \nu 
   n+6 Q^2 \omega
   ^2\nonumber\\
   &+\frac{8 Q^4 \omega ^4}{(\nu
   +n) (\nu +n+1)}\,,\nonumber\\
   &\gamma_n=\frac{i\omega(n+\nu-2iQ\omega)^2}{(n+\nu)(2n+2\nu-1)}\,.
\end{align}
The previous three-terms relation has as consistency relation the following continuous fraction 
\begin{align}
  \beta
   _0-\frac{\alpha _{-1} \gamma
   _0}{\beta _{-1}-\frac{\alpha
   _{-2} \gamma _{-1}}{\beta
   _{-2}-\frac{\alpha _{-3}
   \gamma _{-2}}{\beta
   _{-3}-\dots}}}-\frac{\alpha _0
   \gamma _1}{\beta
   _1-\frac{\alpha _1 \gamma
   _2}{\beta _2-\frac{\alpha _2
   \gamma _3}{\beta _3-\dots}}}=0\,.
\end{align}
The previous continuous fraction can be solved order by order in small-$Q\omega$. 
For the first angular momenta we have relation between the Floquet exponent and the RAM \eqref{RAM}
\beq
\nu^{(\ell)}=k^{(\ell)}-\frac{1}{2} \, , 
\eeq
already discussed extensively and proved in Refs. \cite{Bianchi:2024vmi,Bini:2025ltr,Bini:2025bll}, given the identification of $k$ with the fundamental gauge period $a/\hbar$ of quantum Seiberg-Witten theory. 

The final expression for the incoming radial function is then
\bea
&R_{\rm in}(z)=e^{i\omega z}e^{\frac{i\omega Q^2}{z}} \times \nonumber \\
& \sum_{n=-\infty}^\infty c_n  
z^{\nu+n}U\Big({-}\nu{-}n{-}2iQ\omega,{-}2\nu{-}2n,{-}\frac{2iQ^2\omega}{z}\Big)\,. \nonumber \\
\label {RGiorgio}
\eea
The leading behaviour of the In solution at the horizon is then
\beq
R_{\rm in}(r)\underset{r\to Q}{\simeq}B_{\rm trans}e^{\frac{iQ^2\omega}{r-Q}}(r-Q)^{-2iQ\omega}\,.
\eeq
where 
\beq
B_{\rm trans}=(-2iQ^2\omega)^{\nu+2iQ\omega}\sum_{n=-\infty}^\infty c_n(-2iQ^2\omega)^n
\label {Btrans}
\eeq
is obviously related to the analogous constant ${\cal N}_{in}$ introduced in (\ref {Rpsi})
\beq
B_{\rm trans}={\cal N}_{in}Q^{2iQ\omega} \left ( -\frac {1}{2iQ\omega}  \right )^{\frac {1}{2}-2iQ\omega}\,,
\eeq
which then is expressed in terms of $c_n$ as follows
\beq
{\cal N}_{in}=Q^{-\frac {1}{2}}\sum _{n=-\infty}^{+\infty}c_n (-2i\omega Q^2)^{\frac {1}{2}+\nu+n}\,.
\eeq
The first terms in the PN expansion of $B_{\rm trans}$ are
\bea
B_{\rm trans}^{\ell=0}&=&1+\Big(\pi  Q \omega +2 i Q \omega  \log
   \left(2 Q \omega \right)\Big)\eta^3\nonumber\\
 &+&\frac{1}{2} \left(3
   \sqrt{5}-7\right)\eta^5\nonumber\\
&+&\Big(\frac{1}{2} \pi  \left(\pi +2 i
   \sqrt{5}\right) Q^2 \omega
   ^2-2 Q^2 \omega ^2 \log
   ^2\left(2 Q \omega \right)\nonumber\\
   &-&2
   \left(\sqrt{5}-i \pi \right)
   Q^2 \omega ^2 \log \left(2 Q
   \omega \right)\Big)\eta^6+O\left(\eta^7\right)\,,\nonumber\\
B_{\rm trans}^{\ell=1}&=&-2iQ^2\omega \eta^5\nonumber\\
&+&\Big(-2i\pi Q^3\omega^2+4Q^3\omega^2\log(2Q\omega)\Big)\eta^8\nonumber\\
&+&O\left(\eta^9\right)\,.
\eea
Comparing (\ref {RpsiLess2}) and (\ref {RGiorgio}) we see that,  apart from some constant prefactors, there is the important mismatch due to the overall factor $e^{i\omega z}=\sum_h (Q\omega)^{2h}e^{h(y-\theta)}/h!$,  which is reabsorbed by means of the shift identities \eqref{shift-1}, \eqref{shift-2}. This comparison gives the relation
\begin{widetext}
\bea
\sum _{m'=-\infty} ^{+\infty}c_{m-m'}\,(-i\omega Q^2)^{m-m'}
\sum _{h=|m'|}^{+\infty}\frac {(\omega Q)^{2h}}{h!}\alpha _{h,m'}(k+m-m') =\frac {C_m(k, Q\omega )\sum\limits_{n=-\infty}^{+\infty}c_n (-2i\omega Q^2)^n}{1+\sum\limits_{n=1}^{+\infty}(s_{n,0}+\hat s_{n,1})(-i\omega Q)^{2n}} \,.
\label{checkCc}
\eea
\end{widetext}
We checked  \eqref{checkCc} for $m=0$ expanding both sides in series of $Q \omega$ up to $O((Q \omega)^{8})$. In addition for $m \neq 0$,
calling $T_{m}$ the left hand side of \eqref{checkCc}, we can check the following easier identity:
\bea
T_{m,0}\equiv \frac{T_{m}}{T_{0}}=\frac{C_m(k, Q\omega )}{C_0(k, Q\omega )} \equiv C_{m,0}\,.
\eea
Again this check is done by expanding both sides in series of $Q\omega$ and we did it for the first cases $m=\pm 1,\pm 2, \pm 3$ up to $O ((Q \omega)^{8})$. Here we show the first orders of these expansions (up to $O ((Q \omega)^{5})$ to avoid too lengthy formul{\ae}):
\bea
T_{1,0}&=&C_{1,0} = \frac{i Q^3 \omega ^3}{2
   k^2+k}+\frac{4 Q^4 \omega ^4}{k
   (2 k+1)^2}\nonumber \\
  &+& \frac{4 i Q^5 \omega
   ^5}{k \left(8 k^3+12 k^2-2
   k-3\right)}+O\left((Q\omega)
   ^6\right)\,, \nonumber  \\
   T_{-1,0}&=&C_{-1,0}=\frac{4 i Q^3 \omega ^3}{k (2
   k-1)}-\frac{16 Q^4 \omega ^4}{k
   (2 k-1)^2}\nonumber  \\
   &+&\frac{16 i Q^5
   \omega ^5}{k \left(8 k^3-12
   k^2-2 k+3\right)}+O\left((Q\omega)
   ^6\right) \,,\nonumber \\
   T_{2,0}&=&C_{2,0}=\frac{Q^4 \omega ^4}{64 k
   (k+1)^2}\nonumber \\
   &-&\frac{i Q^5 \omega
   ^5}{4 k \left(4 k^3+12 k^2+11
   k+3\right)}+O\left((Q\omega)
   ^6\right)\,, \nonumber \\
   T_{-2,0} &=&C_{-2,0}= -\frac{Q^4 \omega ^4}{4
   \left((k-1)^2 k\right)}\nonumber \\
  &{-}&\frac{4
   i Q^5 \omega ^5}{k \left(4
   k^3{-}12 k^2{+}11
   k{-}3\right)}+O\left((Q\omega)^{6}\right)  \,.
\eea
However $T_{\pm 3,0}$ and $C_{\pm 3,0}$ are not displayed here because their leading term appears at higher powers of $(Q\omega)^{6}$.\\


%
%

\subsection{MST: Up solution}\label{Upsec}

In the same fashion as for the $R_{\rm in}$, we can write the outgoing solution as
\beq
R_{\rm up}(r)=e^{i\omega z+\frac{iQ^2\omega}{z}}g(z)\,.
\label{RupMST}
\eeq
Putting it in the wave equation \eqref{eqR} we get
\bea
&&0=z^2g''(z)+(2z-2iQ^2\omega+2i\omega z^2)g'(z)\nonumber\\
&&\Big[{-}\ell(\ell{+}1){+}8Q^2\omega^2{+}\frac{4Q^3\omega^2}{z}{+}2\omega z(i{+}2Q\omega)\Big]g(z)\,.\nonumber\\
\eea
This equation can be recast in the form
\bea
&&z^2g''(z)+2z(1+i\omega z)g'(z)+\nonumber\\
&&+\Big[-\nu(\nu+1)+2\omega z(i+2Q\omega)\Big]g(z)\nonumber\\
&&=2iQ^2\omega g'(z)+\Big[\ell(\ell+1)-\nu(\nu+1)\nonumber\\
&&-8Q^2\omega^2-\frac{4Q^3\omega^2}{z}\Big]g(z)\,.
\eea
The differential operator on the left-hand side admits an exact solution in terms of confluent hypergeometric functions:
\beq
g_{\nu}(z)= z^{\nu} U\Big(1{+}\nu{-}2iQ\omega,2{+}2\nu,{-}2iz\omega\Big)\\ \, .
\eeq
Motivated by this solution, we seek the Up solution in the form of the series
\begin{equation}
    g(z)=\sum_{n=-\infty}^\infty c_n \kappa_{n} g_{n+\nu}(z)\,.
\end{equation}
The normalization coefficients \(\kappa_n\) are chosen so that the expansion coefficients \(c_n\) coincide with those appearing in the In solution. Moreover, the RAM parameter \(\nu\) is the same for both the In and Up solutions. Explicitly, we find
\beq
\kappa_n=(-4)^n Q^{2n}\omega^{2n}\,.
\eeq
The explicit expressions for the In and Up solutions in the PN expansion are provided in Appendix \ref{appaA}. 
\\
\\
Finally, let us note that the CT symmetry $z\to Q^2/z$, $\nu\to-\nu-1$ provides a direct map between the In and Up solutions. Making the $\nu$-dependence explicit, it acts on the function  $f(z)$ in the In solution \eqref{RinMST} as
\bea
&f(z;\nu) \overset{CT}{\longmapsto} f\!\left(\frac{Q^{2}}{z};-\nu-1\right)
\nonumber \\
&= Q^{-2\nu}\frac {c_0(-\nu-1)}{c_0(\nu)}g(z;\nu)\,,
\eea
with $g(z)$ appearing in the Up solution \eqref{RupMST}. 
This simple connection between the functions $f(z)$ and $g(z)$ under CT symmetry justifies a posteriori the presence of the $1/z$ factor (which in principle one could think to reabsorbe in the definition of $f(z)$) in (\ref {RinMST}). As anticipated in Subsec. \ref {DCHEwave}, the CT symmetry is equivalent to the symmetry \eqref{psiCT} introduced in integrability variables: indeed, \eqref{psiCT} in gravity variables reads
\beq
\psi _{\rm up}(z;\nu)=\psi _{\rm in}\left (\frac{Q^{2}}{z};-\nu-1\right) \, , 
\eeq
since $q_{1}=q_{2}= -2 i Q \omega$.
\section{Scalar source}
\label{scalar_source}
In eRN spacetime, we consider the wave equation for a massless scalar field, minimally coupled to the background geometry and sourced by a scalar charge density $\rho$ associated with a massive particle moving along a worldline parametrized by $x^\mu=x_p^\mu(\tau)$. The inhomogeneous scalar wave equation can be written as
\begin{equation}
\Box \Phi=-4\pi\rho\,,
\end{equation}
where
\begin{equation}
    \rho=q\int \frac{d\tau}{\sqrt{-g}}\delta^{(4)}(x-x_p(\tau))\,.
\end{equation}
Here, $q$ denotes the scalar charge of the particle. The conditions for circular geodesics, given in \eqref{circorb}, can be solved to obtain the specific energy and angular momentum:
\begin{equation}\label{EJcrit}
    \hat E=\frac{(r-Q)^{3/2}}{r\sqrt{r-2Q}},\quad \hat J=\frac{r}{\sqrt{\frac{r}{Q}-2}}\,,
\end{equation}
where $\hat E=E/\mu$ and $\hat J=J/\mu$.  Integrating the equations of motion in \eqref{4vel} and evaluating the resulting expressions along the circular orbit characterized by \eqref{EJcrit}, we obtain
\begin{align}
    &t=t_p(\tau)=\Gamma \tau,\quad r=r_p(\tau)=r_0,\nonumber\\
    &\theta=\theta_p(\tau)=\frac{\pi}{2},\phi=\phi_p(\tau)=\Gamma\Omega\tau\,.
\end{align}
Here,
\begin{align}
    \Gamma=\frac{ r_0}{\sqrt{(r_0-Q)(r_0-2Q)}}\,,\quad \Omega=\frac{\sqrt{Q}\sqrt{r_0-Q}}{r_0^2}\,.
\end{align}
The source term can be written more explicitly as
\begin{equation}
    \rho=q\int\frac{d\tau}{r^2\sin\theta}\delta^{(4)}(x-x_p(\tau))\,.
\end{equation}
Using the expressions for the particle's worldline along the circular orbit, the four-dimensional delta-function can be decomposed as
\beq
  \delta^{(4)}(x{-}x_p(\tau)){=}\delta(t{-}\Gamma \tau)\delta(r{-}r_0)\delta\left(\theta{-}\frac{\pi}{2}\right)\delta(\phi{-}\Omega t)\,.
\eeq
The integral over $\tau$ can be evaluated straightforwardly
\begin{equation}
    \rho=\frac{q}{r_0^2\Gamma}\delta(r-r_0)\delta\left(\theta-\frac{\pi}{2}\right)\delta(\phi-\Omega t)\,.
\end{equation}
The completeness relation for the spherical harmonics
\begin{equation}
    \delta\left(\theta-\frac{\pi}{2}\right)\delta(\phi-\phi_0)=\sum_{\ell m}Y_{\ell m}(\theta,\phi)Y^*_{\ell m}\left(\frac{\pi}{2},\phi_0\right)\,,
\end{equation}
can be used to rewrite the source density as
\begin{align}
 \rho&=\frac{q}{r_0^2\Gamma}\sum_{\ell m}\delta(r-r_0)Y_{\ell m}(\theta,\phi)Y^*_{\ell m}\left(\frac{\pi}{2},\Omega t\right)\nonumber\\
 &=\frac{q}{r_0^2\Gamma}\sum_{\ell m}\delta(r-r_0)Y_{\ell m}(\theta,\phi)Y^*_{\ell m}\left(\frac{\pi}{2},0\right)e^{-im\Omega t}\nonumber\\
 &=\frac{q}{r_0^2\Gamma}\sum_{\ell m}\delta(r-r_0)Y_{\ell m}(\theta,\phi)Y^*_{\ell m}\left(\frac{\pi}{2},0\right)\nonumber\\
 &\times\int\frac{d\omega}{2\pi}e^{-i\omega t} 2\pi \delta(\omega-m\Omega)\,.
\end{align}
Using the ansatz \eqref{anssf}, we have
\begin{widetext}
    \begin{align}
        \Box\Phi&=\sum_{\ell m}\int\frac{d\omega}{2\pi}e^{-i\omega t}Y_{\ell m}(\theta,\phi)\Big[\left(1-\frac{Q}{r}\right)^2R''_{\ell m \omega}(r)+\frac{2(r-Q)}{r^2}R'_{\ell m \omega}(r)+\left(\frac{r^2\omega^2}{(r-Q)^2}-\frac{\ell(\ell+1)}{r^2}\right)R_{\ell m \omega}(r)\Big]\nonumber\\
        &=-4\pi\sum_{\ell m }\int \frac{d\omega}{2\pi}e^{-i\omega t}\hat{S}_{\ell m}(\omega,r)Y_{\ell m}(\theta,\phi)\,,
    \end{align}
\end{widetext}
where we introduced the source in Fourier space 
\begin{equation}
    \hat{S}_{\ell m}(\omega,r)=\frac{q}{r_0^2\Gamma}\delta(r-r_0)2\pi\delta(\omega-m\Omega)Y^*_{\ell m }\left(\frac{\pi}{2},0\right)\,.
\end{equation}
In short, the inhomogeneous radial equation can be written as
\begin{equation}
    \mathcal{L}_r R_{\ell m \omega}(r)=S_{\ell m \omega}\delta(r-r_0)\,,
\end{equation}
where
\begin{align}\label{source1}
    \mathcal{L}_r&{=}\frac{d^2}{dr^2}{+}\frac{2}{r{-}Q}\frac{d}{dr}{+}\frac{r^4\omega^2}{(r{-}Q)^4}{-}\frac{\ell(\ell{+}1)}{(r{-}Q)^2}\,,\nonumber\\
    S_{\ell m \omega}&=2\pi \delta(\omega-m\Omega)\bar S_{\ell m \omega}\,,\nonumber\\
    \bar S_{\ell m \omega}&=-\frac{4\pi q\sqrt{r_0-2Q}}{(r_0-Q)^\frac{3}{2}r_0}Y^*_{\ell m}\left(\frac{\pi}{2},0\right)\,.
\end{align}
Here, we work in units in which the probe mass $\mu$ is set to unity.

Defining
\begin{equation}
    \Delta(r)=(r-Q)^2\,,
\end{equation}
the equation can be solved using the Green function method
\begin{equation}
    \mathcal{L}_rG_{\ell m \omega}(r,r')=\frac{1}{\Delta(r)}\delta(r-r')\,,
\end{equation}
where we introduced the Green function 
\begin{align}
    G_{\ell m \omega}(r,r')&=\frac{1}{W_{\ell m \omega}}\Big[R_{\rm in}(r)R_{\rm up}(r')\Theta(r'-r)\nonumber\\
    &+R_{\rm in}(r')R_{\rm up}(r)\Theta(r-r')\Big]\,,
\end{align}
and the constant Wronskian
\begin{equation}
    W_{\ell m \omega}=\Delta(r)\Big[R_{\rm in}(r)R'_{\rm up}(r)-R_{\rm up}(r)R'_{\rm in}(r)\Big]\,.
\end{equation}
In this language, the particular solution is
\begin{align}\label{Rnonhom}
    R_{\ell m \omega}(r)&=S_{\ell m \omega}\int dr'G_{\ell m \omega}(r,r')\Delta(r')\delta(r'-r_0)\nonumber\\
    &=S_{\ell m \omega}G_{\ell m \omega}(r,r_0)\Delta(r_0)\,.
\end{align}

\section{Radiated energy}
\label{Energy}
In what follows, we investigate the energy loss due to the emission of massless scalar waves. The radiated energy can be computed from the associated energy-momentum tensor, which, for a massless complex scalar field, is given by
\begin{equation}
    8\pi T^{\rm scal}_{\mu\nu}=\partial_\mu \Phi^*\partial_\nu \Phi+\partial_\mu\Phi\partial_\nu\Phi^*-g_{\mu\nu}\partial_\lambda\Phi^*\partial^\lambda\Phi\,.
\end{equation}
The energy flux at infinity per unit time and solid angle is therefore given by
\begin{equation}
    \frac{d^2E^\infty}{dtd\Omega}=\lim_{r\to\infty}\left(r^2{T^{\rm scal\, r}}_t\right)\,.
\end{equation}
For a complex scalar field the energy-momentum tensor is
\begin{equation}
    8\pi T^{\rm scal}_{rt}=\Phi^*_{,r}\Phi_{,t}+\Phi_{,r}\Phi^*_{,t}\,,
\end{equation}
implying 
\begin{equation}
    8\pi {T^{\rm scal\,r}}_t=\frac{\Delta(r)}{r^2}\left(\Phi^*_{,r}\Phi_{,t}+\Phi_{,r}\Phi^*_{,t}\right)\,.
\end{equation}
Suppressing the dependence on the spacetime variables, we have
\begin{align}\label{recfield}
    \Phi&=\sum_{\ell m}Y_{\ell m}\int\frac{d\omega}{2\pi}e^{-i\omega t}R_{\ell m \omega}\,,\nonumber\\
    \Phi_{,t}&=\sum_{\ell m}Y_{\ell m}\int\frac{d\omega}{2\pi}(-i\omega)e^{-i\omega t}R_{\ell m \omega}\,,\nonumber\\
    \Phi_{,r}^*&=\sum_{\ell'm'}Y_{\ell'm'}^*\int\frac{d\omega'}{2\pi}e^{i\omega't}\frac{d}{dr}R^*_{\ell'm'\omega'}\,,
\end{align}
and therefore
\begin{widetext}
    \begin{align}
        r^2{T^{\rm scal\,r}}_t&=\frac{\Delta(r)}{8\pi}\left(\Phi^*_{,r}\Phi_{,t}+\Phi_{,r}\Phi^*_{,t}\right)\nonumber\\
        &=\frac{\Delta(r)}{8\pi}\sum_{\ell m, \ell'm'}\int\frac{d\omega}{2\pi}\int\frac{d\omega'}{2\pi}Y_{\ell m}(\theta,\phi)Y^*_{\ell'm'}(\theta,\phi)\Bigg[(-i\omega)e^{-i(\omega-\omega')t}R_{\ell m \omega}(r)\frac{d}{dr}R^*_{\ell'm'\omega'}(r)\nonumber\\
        &+(i\omega')e^{i(\omega-\omega')t}R^*_{\ell'm'\omega'}(r)\frac{d}{dr}R_{\ell m \omega}(r)\Bigg]\,.
    \end{align}
\end{widetext}
Using the orthogonality properties of the spherical harmonics
\begin{equation}
    \int \sin\theta d\theta d\phi Y^*_{\ell m}(\theta,\phi)Y_{\ell'm'}(\theta,\phi)=\delta_{\ell\ell'}\delta_{mm'}\,,
\end{equation}
we find
\begin{align}
    \frac{dE^\infty}{dt}&=\lim_{r\to\infty}\int \sin\theta d\theta d\phi r^2{T^{\rm scal\, r}}_t\nonumber\\
    &=\lim_{r\to \infty}\frac{\Delta(r)}{8\pi}\sum_{\ell m}\int\frac{d\omega}{2\pi}\int\frac{d\omega'}{2\pi}\nonumber\\
    &\times\Bigg[(-i\omega) e^{-i(\omega-\omega')t}R_{\ell m \omega}(r)\frac{d}{dr}R^*_{\ell m \omega'}(r)\nonumber\\
    &+(i\omega') e^{i(\omega-\omega')t}R^*_{\ell m \omega'}(r)\frac{d}{dr}R_{\ell m \omega}(r)\Bigg]\,.
    \label{dedt}
\end{align}
Because of the $r\to \infty$ limit we can use here only the Up-part of the $R_{\ell m \omega}(r)$ solution 
\begin{align}
    R_{\ell m \omega}&=\mathfrak{R}_{\ell m \omega}(r)2\pi\delta(\omega-m\Omega)\,,\nonumber\\
    \mathfrak{R}_{\ell m \omega}(r)&=\frac{R_{\rm up}(r)R_{\rm in}(r_0)}{W_{\ell m \omega}}\Delta(r_0)\bar S_{\ell m\omega}\,.
    \label{Rlmw}
\end{align}
Finally
\begin{widetext}
    \begin{align}
        \frac{dE^\infty}{dt}&=\lim_{r\to \infty}\frac{\Delta(r)}{8\pi}\sum_{\ell m}(-im\Omega)\Bigg[\mathfrak{R}_{\ell m\omega}(r)\frac{d}{dr}\mathfrak{R}_{\ell m \omega}^*(r)-\mathfrak{R}^*_{\ell m\omega}(r)\frac{d}{dr}\mathfrak{R}_{\ell m \omega}(r)\Bigg]_{\omega=m\Omega}\nonumber\\
        &=\lim_{r\to\infty}\frac{\Delta(r)}{4\pi}\sum_{\ell m}m\Omega\,\mathfrak{I}m\Bigg[\mathfrak{R}_{\ell m\omega}(r)\frac{d}{dr}\mathfrak{R}_{\ell m \omega}^*(r)\Bigg]_{\omega=m\Omega}\,.
    \end{align}
To evaluate the energy flux using the above expression, we employ the homogeneous In and Up solutions for the multipole modes $\ell=0,1,2,3,4,5$. For each aforementioned multipole we find
\bea
\frac{Q^2}{q^2}\frac{dE^\infty_{\ell=0}}{dt}&=&0\,,\nonumber\\
\frac{Q^2}{q^2}\frac{dE^\infty_{\ell=1}}{dt}&{=}&\frac{u^4}{3}{-}\frac{12}{5}u^5{+}\frac{2}{3}\pi u^\frac{11}{2}{+}\frac{3503}{525}u^6{-}\frac{77}{15}\pi u^\frac{13}{2}{-}\left(\frac{4051}{2025}{-}\frac{4}{9}\pi^2{+}\frac{2}{3}\log(4u e^{2\gamma})\right)u^7{+}\frac{10963}{700}\pi u^\frac{15}{2}{+}O\left(u^8\right) \,,\nonumber\\
\frac{Q^2}{q^2}\frac{dE^\infty_{\ell=2}}{dt}&=&\frac{16}{15}u^5-\frac{1184}{105}u^6+\frac{64}{15}\pi u^\frac{13}{2}+\frac{326272}{6615}u^7-\frac{992}{21}\pi u^\frac{15}{2}+O\left(u^8\right)\,,\nonumber\\
\frac{Q^2}{q^2}\frac{dE^\infty_{\ell=3}}{dt}&=&\frac{1367}{525}u^6-\frac{172241}{4725}u^7+\frac{16403}{1050}\pi u^\frac{15}{2}+O\left(u^8\right)\,,\nonumber\\
\frac{Q^2}{q^2}\frac{dE^\infty_{\ell=4}}{dt}&=&\frac{114752}{19845}u^7+O\left(u^8\right)\,,\nonumber\\
\frac{Q^2}{q^2}\frac{dE^\infty_{\ell=5}}{dt}&=&O\left(u^{8}\right)\,,
\eea
where $u=Q/r_0$ and $\eta=1$. Plugging all together, our final result is
\bea\label{dEdtfin}
\frac{Q^2}{q^2}\frac{dE^\infty}{dt}{=}\frac{u^4}{3}{-}\frac{4u^5}{3}{+}\frac{2}{3}\pi u^\frac{11}{2}{-}2u^6{-}\frac{13}{15}\pi u^\frac{13}{2}{+}\left(\frac{2248}{135}{+}\frac{4}{9}\pi^2{-}\frac{2}{3}\log(4ue^{2\gamma})\right)u^7{-}\frac{6701}{420}\pi u^\frac{15}{2}{+}O\left(u^8\right)\,.
\eea
In terms of the variable 
\begin{equation}\label{yu}
\mathcal{Y}=(Q\Omega)^\frac{2}{3}=\left(u^2\sqrt{\frac{1}{u}-1}\right)^\frac{2}{3},\qquad u=\mathcal{Y}+\frac{1}{3}\mathcal{Y}^2+\frac{1}{3}\mathcal{Y}^3+\frac{35}{81}\mathcal{Y}^4+\frac{154}{243}\mathcal{Y}^5+\mathcal{Y}^6+\frac{10868}{6561}\mathcal{Y}^7+O\left(\mathcal{Y}^8\right)\,,
\end{equation}
the flux becomes
\begin{equation}
\frac{Q^2}{q^2}\frac{dE^\infty}{dt}=\frac{\mathcal{Y}^4}{3}-\frac{8}{9}\mathcal{Y}^5+\frac{2}{3}\pi \mathcal{Y}^\frac{11}{2}-\frac{32}{9}\mathcal{Y}^6+\frac{16}{45}\pi \mathcal{Y}^\frac{13}{2}+\mathcal{Y}^7\left(\frac{12172}{1215}+\frac{4}{9}\pi^2-\frac{2}{3}\log(4\mathcal{Y}e^{2\gamma})\right)-\frac{9887}{630}\pi \mathcal{Y}^\frac{15}{2} +O(\mathcal{Y}^{8}) \,.
\end{equation}
This analytical result agrees very well with the numerical values reported in Table I of Ref. \cite{Castillo:2018ibo}, as shown in Fig.~\ref{figdedt}.
\begin{figure}
\centering
\begin{minipage}[c]{0.48\linewidth}
    \centering
    \includegraphics[width=\linewidth]{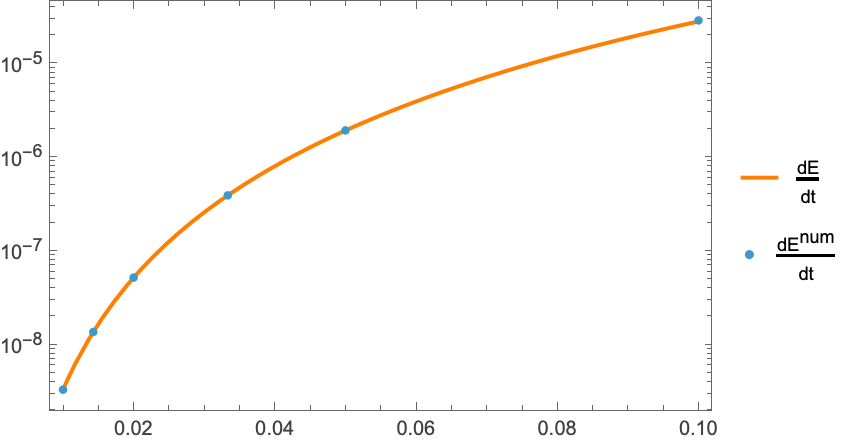}
\end{minipage}
\hfill
\begin{minipage}[c]{0.48\linewidth}
    \centering
    \renewcommand{\arraystretch}{1.3}
    \small
    \begin{tabular}{|c|c|c|c|}
    \hline
    $\mathcal{Y}$ & $\frac{dE^{\rm num}}{dt}$ & $\frac{dE}{dt}$ & {\rm Error(\%)} \\
    \hline
    $\frac{1}{10}$  & $2.8436184\times10^{-5}$ & $2.7730662\times10^{-5}$ & $2.5$   \\
    \hline
    $\frac{1}{20}$  & $1.9061791\times10^{-6}$ & $1.9031307\times10^{-6}$  & $0.16$ \\
    \hline
    $\frac{1}{30}$  & $3.8647892\times10^{-7}$ & $3.8635448\times10^{-7}$ & $0.032$  \\
    \hline
    $\frac{1}{50}$  & $5.1232128\times10^{-8}$ & $5.1229942\times10^{-8}$ & $4.3*10^{-3}$  \\
    \hline
    $\frac{1}{70}$  & $1.3475403\times10^{-8}$ & $1.3475252\times10^{-8}$  & $1.1*10^{-3}$ \\
    \hline
    $\frac{1}{100}$ & $3.2620620\times10^{-9}$ & $3.2620530\times10^{-9}$ & $2.8*10^{-4}$  \\
    \hline
    \end{tabular}
\end{minipage}
\caption{Comparison of the flux of energy at infinity $\frac{dE}{dt}$ in Eq \eqref{dEdtfin} as function of $\mathcal{Y}$ with the numerical data relative to the same quantity (denoted $\frac{dE^{\rm num}}{dt}$ in this plot) appearing in Tab.I of \cite{Castillo:2018ibo}.}\label{figdedt}
\end{figure}
Similarly the flux at the horizon turns out to be
\beq
\frac{dE^H}{dt}=\frac{4\pi q^2 Q^2}{\Gamma^2}\sum_{\ell m}\omega^2\left|Y_{\ell m}\left(\frac{\pi}{2},0\right)\right|^2\frac{|B_{\rm trans}|^2|R_{\rm up}(r_0)|^2}{|W|^2}\bigg|_{\omega=m\Omega}\,,
\eeq
\end{widetext}
where $B_{\rm trans}$ is given by (\ref {Btrans}). 
Eventually, our final expression for the energy flux at the horizon expressed in the dimensionless variable $u=Q/r_0$ reads
\beq\label{dEdtHfin}
\frac{dE^H}{dt}=\frac{q^2}{Q^2}\frac{u^{10}}{3}+O\left(u^{11}\right)\,.
\eeq
\section{Self field}
\label{self_field}
Using the definition in Eq.~\eqref{Rnonhom} together with the source term in Eq.~\eqref{source1}, we obtain the following explicit expression for the scalar field
\bea
\Phi(t,r,\theta,\phi)
&=&
-\frac{4\pi q}{r_0}
\sqrt{(r_0-2Q)(r_0-Q)}\nonumber\\
&\times&
\sum_{\ell m}
Y_{\ell m}\left(\theta,\phi-\Omega t\right)
Y^*_{\ell m}\left(\frac{\pi}{2},0\right)\nonumber\\
&\times&
\left.
\frac{R_{\rm in}(r_0)R_{\rm up}(r)}{W}
\right|_{\omega=m\Omega}\, , \nonumber\\
\eea
which can then be evaluated along the particle's geodesic
\bea\label{phiorb}
&&\Phi\left(t,r_0,\frac{\pi}{2},\Omega t\right)
=
-\frac{4\pi q}{r_0}
\sqrt{(r_0-2Q)(r_0-Q)}\nonumber\\
&\times&
\sum_{\ell m}
\left|Y_{\ell m}\left(\frac{\pi}{2},0\right)\right|^2\left.
\frac{R_{\rm in}(r_0)R_{\rm up}(r_0)}{W}
\right|_{\omega=m\Omega}\,.\nonumber\\
\eea

\subsection{PN contribution}
Here, we compute the self-field \eqref{phiorb} using the PN-expanded solutions presented in Appendix \ref{appaB}. After evaluating the Green function at $\omega=m\Omega$, the sum over $m$, 
\beq
Y_{\ell}(d)=\sum_{m=-\ell}^\ell \left|Y_{\ell m}\left(\frac{\pi}{2},0\right)\right|^2 m^{d}\,,
\eeq
can be done using the following identity
\bea
\Gamma_{\ell}(z)&=&\sum_{m=-\ell}^\ell e^{m z}\left|Y_{\ell m}\left(\frac{\pi}{2},0\right)\right|^2\nonumber\\
&=&\frac{2\ell+1}{4\pi}e^{\ell z}{}_2F_1\left(\frac{1}{2},-\ell,1,1-e^{-2z}\right)\nonumber\\
&=&Y_\ell(0)+\frac{z^2}{2!}Y_\ell(2)+\dots \, , 
\eea
i.e. explicitly
\bea
Y_\ell(0)&=&\frac{1+2\ell}{4\pi}\,,\nonumber\\
Y_\ell(2)&=&\frac{(1+2\ell)}{4\pi}\frac{\ell(\ell+1)}{2}\,,\nonumber\\
Y_\ell(4)&=&\frac{(1+2\ell)}{4\pi}\frac{\ell(\ell+1)(3\ell^2+3\ell-2)}{8}\,,\nonumber\\
Y_\ell(6)&=&\frac{(1+2\ell)}{4\pi}\frac{\ell(\ell+1)(5\ell^4+10\ell^3-5\ell^2-10\ell+8)}{16}\,,\nonumber\\
Y_\ell(8)&=&\frac{(1+2\ell)}{4\pi}\frac{\ell(\ell+1)}{128}(35\ell^6+105\ell^5-35\ell^4\nonumber\\
&-&245\ell^3+168\ell^2+308\ell-272)\,.
\eea

From \eqref{phiorb}, the total field is
\beq
\Phi_{\rm orb}=\Phi\left(t,r_0,\frac{\pi}{2},\Omega t\right)=\sum_{\ell=0}^\infty \Phi_\ell \, , 
\eeq
where
\begin{widetext}
    \bea
\Phi_\ell&=&\frac{q}{r_0}\Bigg[1-\frac{Q \left(2 \ell ^2+2 \ell
   -3\right)}{2 r_0 (2 \ell
   -1) (2 \ell +3)}\eta^2-\frac{Q^2 \left(46 \ell ^4+92
   \ell ^3-146 \ell ^2-192 \ell
   -135\right)}{8 r_0^2 (2
   \ell -3) (2 \ell -1) (2 \ell
   +3) (2 \ell +5)}\eta^4\nonumber\\
   &-&\frac{Q^3 \left(2192 \ell
   ^{10}{+}10960 \ell ^9{-}10612 \ell
   ^8{-}108208 \ell ^7{-}55486 \ell
   ^6{+}258302 \ell ^5{+}72316 \ell
   ^4{-}434034 \ell ^3{-}223765 \ell
   ^2{+}57625 \ell
   +158550\right)}{16 r_0^3
   (\ell -1) (\ell +2) (2 \ell
   -5) (2 \ell -3) (2 \ell -1) (2
   \ell +1)^2 (2 \ell +3) (2 \ell
   +5) (2 \ell +7)}\eta^6\nonumber\\
   &-&\frac{\eta^8}{128 r_0^4
   (\ell -2) (\ell -1) (\ell +2)
   (\ell +3) (2 \ell -7) (2 \ell
   -5) (2 \ell -3) (2 \ell -1)^2
   (2 \ell +1)^2 (2 \ell +3)^2 (2
   \ell +5) (2 \ell +7) (2 \ell
   +9)}\nonumber\\
   &\times&\Bigg(Q^4 \left(431840 \ell
   ^{16}+3454720 \ell
   ^{15}-3941488 \ell
   ^{14}-88048016 \ell
   ^{13}-93804730 \ell
   ^{12}+738985588 \ell
   ^{11}+1331726978 \ell
   ^{10}\right.\nonumber\\
   &-&\left.2625564848 \ell
   ^9-7043577122 \ell ^8+46349876
   \ell ^7+11130261886 \ell
   ^6+9509138576 \ell
   ^5+2774854877 \ell
   ^4+1376346346 \ell
   ^3\right.\nonumber\\
   &-&\left.1819131831 \ell
   ^2-2683841832 \ell
   +1097904780\right)\Bigg)\Bigg]+O\left(\eta^{10}\right)\,.
    \eea
\end{widetext}
However, the mode sum diverges when evaluated on the particle's worldline. We therefore need to subtract the singular field, which is given by
\bea
B&=&\frac{q}{r_0}\Bigg(1{-}\frac{Q}{4r_0}\eta^2{-}\frac{23Q^2}{64r_0^2}\eta^4{-}\frac{137Q^3}{256r_0^3}\eta^6{-}\frac{13495Q^4}{16384r_0^4}\eta^8\nonumber\\
&-&\frac{85601Q^5}{65536r_0^5}\eta^{10}+O\left(\eta^{11}\right)\Bigg)\,.
\eea
The regularized field is obtained by subtracting the singular field from the full field
\bea
\Phi^{\ell}_{\rm orb, reg}=\Phi_\ell-B\,.
\eea
However, the resulting quantity remains divergent for $\ell=0,1,2,3$. The corresponding singular behaviors for these multipoles are given by
\bea
\Phi^{\ell=0}_{\rm orb, reg}&\underset{\ell\to 0}{\sim}&-\frac{3qQ^5\eta^{10}}{r_0^6\ell}\,,\nonumber\\
\Phi^{\ell=1}_{\rm orb, reg}&\underset{\ell\to 1}{\sim}&\frac{q}{\ell-1}\Bigg(\frac{2Q^3\eta^6}{3r_0^4}-\frac{19Q^4\eta^8}{5r_0^5}+\frac{16229Q^5\eta^{10}}{2100r_0^6}\Bigg)\,,\nonumber\\
\Phi^{\ell=2}_{\rm orb, reg}&\underset{\ell\to 2}{\sim}&\frac{q}{\ell-2}\Bigg(\frac{32Q^4\eta^8}{15r_0^5}-\frac{2032Q^5\eta^{10}}{105r_0^6}\Bigg)\,,\nonumber\\
\Phi^{\ell=3}_{\rm orb, reg}&\underset{\ell\to 3}{\sim}&\frac{q}{\ell-3}\frac{2734Q^5\eta^{10}}{525r_0^6}\,.
\eea
To achieve an accuracy through $\mathcal{O}(\eta^{10})$, we evaluate the PN mode sum for $\ell\geq 4$, while the missing contributions from $\ell=0,1,2,3$ will be computed separately and added to the result using either the MST-like method or the kink approach, which provide equivalent formulations. The contribution obtained from the PN solution is therefore
\bea
\Phi^{PN}_{\rm orb, reg}&=&\sum_{\ell=4}^\infty \Phi^{\ell}_{\rm orb, reg}\nonumber\\
&=&\frac{q}{Q}\Bigg[\frac{u^2}{21}-\frac{73 u^3}{3696}+\left(\frac{563008939}{302702400}
   -\frac{3 \pi ^2}{16}\right)
   u^4\nonumber\\
   &+&\left(\frac{15 \pi
   ^2}{64}-\frac{227852437}{11070
   2592}\right) u^5\nonumber\\
   &+&\left(-\frac{18792665465071}{1449
   0969292800}-\frac{411 \pi
   ^2}{1024}\right) u^6+O\left(u^7\right)\Bigg]\,,\nonumber\\
\eea
where, we have introduced the dimensionless variable \(u=Q/r_0\) and set \(\eta=1\).

The contributions from $\ell=0,1,2,3$ are
\bea
\Phi^{\ell=0}_{\rm orb, reg}&=&-\frac{q}{Q}\Bigg[\frac{u^2}{4}{+}\frac{17 u^3}{64}{+}\frac{71 u^4}{256}{+}\frac{4553 u^5}{16384}{+}\frac{16543 u^6}{65536}\Bigg]\nonumber\\
&+&O\left(u^7\right)\,,\nonumber\\
\Phi^{\ell=1}_{\rm orb, reg}&=&\frac{q}{Q}\Bigg[\frac{3 u^2}{20}-\frac{375 u^3}{448}+u^4 \left(-\frac{2}{3} \log (4
   ue^{2\gamma})\right.\nonumber\\
   &+&\left.\frac{660521}{241920}\right)+u^5 \left(\frac{19}{5} \log (4
   u e^{2\gamma})\right.\nonumber\\
   &-&\left.\frac{4505808767}{8515584
   00}\right)-\frac{2}{3} \pi  u^{11/2}\nonumber\\
   &+&u^6 \left(-\frac{16229 \log (4
   ue^{2\gamma})}{2100}-\frac{14609007485231}{
   1549836288000}\right)\Bigg]\nonumber\\
   &+&O\left(u^7\right)\,,\nonumber\\
\Phi^{\ell=2}_{\rm orb, reg}&=&\frac{q}{Q}\Bigg[\frac{u^2}{28}+\frac{155 u^3}{1344}-\frac{6870587 u^4}{4435200}\nonumber\\
&+&u^5 \left(-\frac{32}{15} \log (16
   ue^{2\gamma})+\frac{931740638683}{7749
   1814400}\right)\nonumber\\
   &+&u^6 \left(\frac{2032}{105} \log
   \left(16 u e^{2\gamma}\right)-\frac{3245104626823}{61
   993451520}\right)\Bigg]\nonumber\\
   &+&O\left(u^7\right)\,,\nonumber\\
\Phi^{\ell=3}_{\rm orb, reg}&=&\frac{q}{Q}\Bigg[\frac{u^2}{60}+\frac{5 u^3}{704}+\frac{4859977 u^4}{26906880}-\frac{228533021221
   u^5}{77491814400}\nonumber\\
   &{+}&u^6 \left({-}\frac{2734}{525} \log
   (4 ue^{2\gamma}){+}\frac{1446260281189181}
   {41402769408000}\right.\nonumber\\
   &-&\left.\frac{729
   \log (3)}{70}\right)\Bigg]+O\left(u^7\right)\,.\nonumber\\
\eea
Summing all together, the final result for the self-field on the orbit is
\bea
\Phi_{\rm orb}&{=}&-\frac{q}{Q}\Bigg[u^3+u^4\Big(\frac{3\pi^2}{16}-\frac{53}{18}+\frac{4}{3}\log(2)\nonumber\\
&+&\frac{2}{3}\log(ue^{2\gamma})\Big)+u^5\Big(\frac{14}{15}\log(2)-\frac{521}{360}-\frac{15}{64}\pi^2\nonumber\\
&-&\frac{5}{3}\log(ue^{2\gamma})\Big)+\frac{2}{3}\pi u^\frac{11}{2}+u^6\Big(-\frac{77}{12} \log \left(e^{2
   \gamma } u\right)\nonumber\\
   &+&\frac{411 \pi
   ^2}{1024}+\frac{9539}{336}+\frac{729 \log (3)}{70}-\frac{10823
   \log (2)}{210}\Big)\Bigg]\nonumber\\
   &+&O\Big(u^7\Big)\,.
\eea
In terms of the variable $\mathcal{Y}$ in \eqref{yu}, the self-field is
\bea
\Phi_{\rm orb}&{=}& -\frac{q}{Q} \bigg[\mathcal{Y}^3+\mathcal{Y}^4\left(-\frac{35}{18}+\frac{3\pi^2}{16}+\frac{2}{3}\log(4\mathcal{Y}e^{2\gamma})\right)\nonumber\\
&+&\mathcal{Y}^5\Big(-\frac{4123}{1080}+\frac{\pi ^2}{64}+\frac{64 \log (2)}{15}\nonumber\\
&-&\frac{7}{9} \log \left(4 e^{2
   \gamma } \mathcal{Y}\right)\Big)+\mathcal{Y}^6\Big(\frac{7397}{336}+\frac{395 \pi ^2}{1024}\nonumber\\
   &-&\frac{9952 \log (2)}{315}+\frac{729 \log (3)}{70}\nonumber\\
   &-&\frac{283}{36} \log \left(4 e^{2
   \gamma } \mathcal{Y}\right)\Big)+\frac{2}{3}\pi \mathcal{Y}^\frac{11}{2}+O\left(\mathcal{Y}^7\right) \bigg]\,.
\eea
\section{Scalar self force}
\label{scalar_self_force}
The scalar self-force describes the backreaction of the particle's own regular scalar field on its motion. Since it is gauge invariant in the present framework, it represents a physically meaningful quantity that can be related to observables. Its definition is
\beq
F_\alpha(x^\mu)=q\nabla_\alpha \Phi(x^\mu)\,.
\eeq
After summing over $m$, the divergent part for large $\ell$ is removed by
\beq
F_\alpha^{\rm reg}=\sum_{\ell}\Big[\frac{1}{2}\Big(F^\ell_{\alpha,R}+F^\ell_{\alpha,L}\Big)-B_\alpha\Big]\,,
\eeq
where $F^\ell_{\alpha,R,L}$ denote the limits $r\to r_0^\pm$ of each mode and $B_\alpha$ are regularization parameters, whose expressions are
\bea
B_t&=&B_\phi=0\nonumber\\
B_r&=&-\frac{q^2}{131072 r_0^7}\Big[65536 r_0^5+16384 \eta ^2 Q r_0^4\nonumber\\
&+&5120 \eta ^4 Q^2 r_0^3
-13056 \eta ^6 Q^3 r_0^2-43004 \eta ^8 Q^4 r_0\nonumber\\
&-&93271 \eta ^{10} Q^5\Big]+O\left(\eta^{11}\right)\,.
\eea
The regularized components of the self-force which consider contributions of $\ell=0,1,2,3$ of MST-like solutions are
\begin{widetext}
    \bea\label{Ftdiu}
\frac{Q^2}{q^2}F_t&{=}&\frac{u^4}{3}-\frac{5}{6}u^5+\frac{2\pi}{3}u^\frac{11}{2}-\frac{77}{24}u^6+\frac{2\pi}{15}u^\frac{13}{2}+u^7\left(\frac{25483}{2160}-\frac{2}{3}\log (4ue^{2\gamma})+\frac{4 \pi
   ^2}{9}\right)-\frac{1097\pi}{70}u^\frac{15}{2}+O\left(u^8\right)\,,\nonumber\\
\frac{Q^2}{q^2}F_r&{=}&u^5\left(\frac{3 \pi
   ^2}{32}-\frac{2}{9}-\frac{2}{3} \log (4ue^{2\gamma})\right)+u^6\left(-\frac{9 \pi ^2}{128}-2
   \gamma
   +\frac{203}{15}-\frac{128}{15}\log(2)-\log(4ue^{2\gamma})\right)\nonumber\\
   &-&\frac{2\pi}{3}u^\frac{13}{2}+u^7\left(\frac{1263 \pi
   ^2}{2048}-\frac{30487}{1260}+\frac{2496}{35}\log(2)-\frac{2187}{70}\log(3)+\frac{181}{12}\log(4ue^{2\gamma})\right)+O\left(u^8\right)\,,\nonumber\\
   F_\phi&=&-\frac{F_t}{\Omega}\,.
    \eea
where for notational simplicity, we henceforth omit the superscript \lq\lq reg". Here, we set $\eta=1$ and introduce the dimensionless variable $u=Q/r_0$. 

As a consistency check using \eqref{Ftdiu}, \eqref{dEdtfin} and \eqref{dEdtHfin}, the following identity involving the dissipative component of the self-force $F_t$ is satisfied
\begin{equation}
F^t=\Gamma\left(\frac{dE^\infty}{dt}+\frac{dE^H}{dt}\right)\,.
\end{equation}
In terms of the variable $\mathcal{Y}$ in \eqref{yu}, we have
  \bea\label{FtFr}
  \frac{Q^2}{q^2}F_t&{=}&\frac{\mathcal{Y}^4}{3}-\frac{7 \mathcal{Y}^5}{18}+\frac{2}{3} \pi  \mathcal{Y}^{11/2}-\frac{283 \mathcal{Y}^6}{72}+\frac{61}{45} \pi  \mathcal{Y}^{13/2}+\mathcal{Y}^7\Big[\frac{80407}{19440}+\frac{4 \pi ^2}{9}-\frac{2}{3}\log(4\mathcal{Y} e^{2\gamma})\Big]-\frac{16687 \pi  \mathcal{Y}^{15/2}}{1260}+O\left(\mathcal{Y}^8\right)\,,\qquad\nonumber\\
   \frac{Q^2}{q^2}F_r&{=}&\mathcal{Y}^5\Big[-\frac{2}{9}+\frac{3 \pi ^2}{32}-\frac{2}{3}\log(4\mathcal{Y} e^{2\gamma})\Big]+\mathcal{Y}^6\Big[\frac{1747}{135}+\frac{11 \pi ^2}{128}-\frac{128 \log (2)}{15}-\frac{19}{9} \log \left(4 e^{2
   \gamma } \mathcal{Y}\right)\Big]-\frac{2}{3} \pi  \mathcal{Y}^{13/2}\nonumber\\
   &+&\mathcal{Y}^7\Big[\frac{15473}{11340}+\frac{4525 \pi ^2}{6144}+\frac{5696 \log (2)}{105}-\frac{2187 \log (3)}{70}+\frac{1213}{108} \log \left(4 e^{2
   \gamma } \mathcal{Y}\right)\Big]+O\left(\mathcal{Y}^8\right)\,.
    \eea
The analytical results for $F_t$ and $F_r$ agree with the numerical values reported in Tables IV and V, respectively, of Ref. \cite{Castillo:2018ibo}. These comparisons are shown in Figs.~\ref{figft} and \ref{Frfigr}.



 \begin{figure}
\centering
\begin{minipage}[c]{0.5\linewidth}
    \centering
    \includegraphics[width=\linewidth]{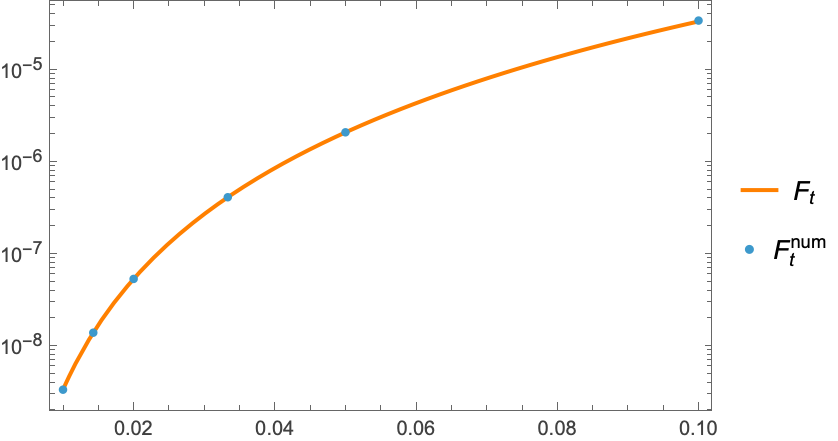}
\end{minipage}
\hfill
\begin{minipage}[c]{0.48\linewidth}
    \centering
    \renewcommand{\arraystretch}{1.3}
    \small
    \begin{tabular}{|c|c|c|c|}
    \hline
    $\mathcal{Y}$ & $F_t^{\rm num}$ & $F_t$ & ${\rm Error(\%)}$\\
    \hline
    $\frac{1}{10}$  & $3.3739003\times 10^{-5}$ & $3.3004323\times 10^{-5}$ & $2.2$ \\
    \hline
    $\frac{1}{20}$  & $2.0644552\times 10^{-6}$ & $2.0612469\times 10^{-6}$ & $0.16$ \\
    \hline
    $\frac{1}{30}$  & $4.0712045\times 10^{-7}$ & $4.0699831\times 10^{-7}$ & $0.030$ \\
    \hline
    $\frac{1}{50}$  & $5.2830590\times 10^{-8}$ & $5.2828240\times 10^{-8}$ & $4.4*10^{-3}$ \\
    \hline
    $\frac{1}{70}$  & $1.3772312\times 10^{-8}$ & $1.3772148\times 10^{-8}$ & $1.2*10^{-3}$ \\
    \hline
    $\frac{1}{100}$ & $3.3119508\times 10^{-9}$ & $3.3119411\times 10^{-9}$  & $3.0*10^{-4}$\\
    \hline
    \end{tabular}
\end{minipage}
\caption{Comparison of $F_t$ in Eq \eqref{FtFr} as function of $\mathcal{Y}$  with the numerical data relative to the same quantity (denoted $F_t^{\rm num}$ in this plot) appearing in Tab.IV of \cite{Castillo:2018ibo}.}\label{figft}
\end{figure}

\begin{figure}
\centering
\begin{minipage}[c]{0.48\linewidth}
    \centering
    \includegraphics[width=\linewidth]{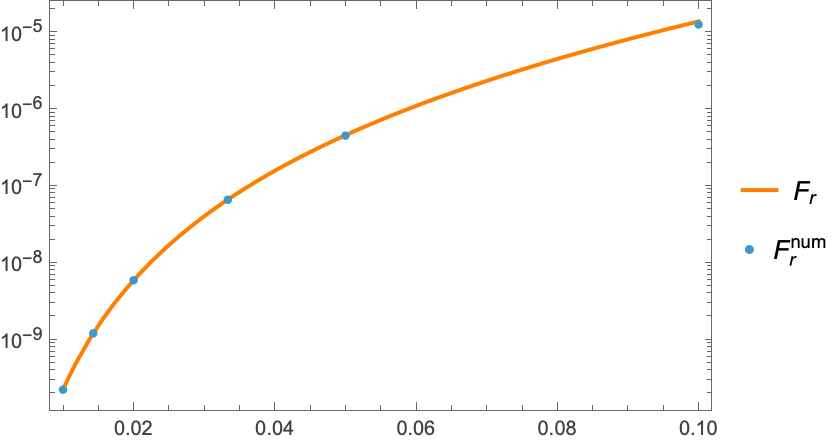}
\end{minipage}
\hfill
\begin{minipage}[c]{0.48\linewidth}
    \centering
    \renewcommand{\arraystretch}{1.3}
    \small
    \begin{tabular}{|c|c|c|c|}
    \hline
    $\mathcal{Y}$ & $F_r^{\rm num}$ & $F_r$ & ${\rm Error(\%)}$\\
    \hline
    $\frac{1}{10}$  & $1.2431961\times10^{-5}$  & $1.3610515\times10^{-5}$ & $9.5$    \\
    \hline
    $\frac{1}{20}$  & $4.4425415\times10^{-7}$  & $4.5053371\times10^{-7}$ & $1.4$   \\
    \hline
    $\frac{1}{30}$  & $6.5123254\times10^{-8}$  & $6.5410349\times10^{-8}$ & $0.44$   \\
    \hline
    $\frac{1}{50}$  & $5.8352086\times10^{-9}$  & $5.8409972\times10^{-9}$ & $0.10$   \\
    \hline
    $\frac{1}{70}$  & $1.1898372\times10^{-9}$  & $1.1902762\times10^{-9}$ & $0.037$   \\
    \hline
    $\frac{1}{100}$ & $2.1981670\times10^{-10}$ & $2.1984508\times10^{-10}$ & $0.013$  \\
    \hline
    \end{tabular}
\end{minipage}
\caption{Comparison of $F_r$ in Eq \eqref{FtFr} as function of $\mathcal{Y}$  with the numerical data relative to the same quantity (denoted $F_r^{\rm num}$ in this plot) appearing in Tab.V of \cite{Castillo:2018ibo}.}
\label{Frfigr}
\end{figure}

\end{widetext}
\section{Concluding remarks}
\label{conclusions}

In this paper we have investigated the emission of massless scalar waves and the scalar self-force experienced by a massive scalar-charged particle moving along circular geodesics in eRN spacetime. Our analysis provides a systematic analytical framework for studying these phenomena within the PN expansion and beyond, exploiting the special mathematical structure of the radial wave equation in the extremal case.

The main difficulty lies in solving the homogeneous radial equation, which takes the form of a DCHE and does not admit a known explicit closed-form solution in general. To overcome this obstacle, we have developed two approaches for constructing its solutions in an exact expansion form. The first, also called kink method, is inspired by quantum integrability theory, while the second is an MST inspired method, suitably adapted to the DCHE: they have both their own advantages. These complementary formulations provide an analytical description of the radial solutions and allow us to systematically extract the PN expansion. Their comparison also establishes an easy relation between the RAM and the Floquet exponent, which appears manifestly in the kink method.

We have applied this framework to compute the energy fluxes carried by scalar waves both to infinity and to the horizon. In addition we computed the regularized scalar field along circular geodesics. The resulting analytical fluxes are in excellent agreement with the numerical data available in literature. We have also obtained PN expressions for the components of the scalar self-force, finding excellent agreement with the corresponding numerical results in \cite{Castillo:2018ibo}.

Compared with the analysis presented in Ref. \cite{Bini:2016egn}, the present work aims to provide a new and complementary methodological framework for studying the scalar self-force in eRN spacetime. 

A further useful feature of the eRN problem is its CT symmetry, which relates the horizon and asymptotic regions of the spacetime. This symmetry is reflected in the structure of the radial solutions and provides an additional consistency check on their analytical construction \eqref{psiCT}.

The methods developed here open several avenues for further investigations. In particular, the kink method can be used as formulated in \cite{ Fioravanti:2025bts} to more general orbital configurations and to investigate higher-order PN contributions to radiative observables and the scalar self-force. More broadly, the agreement between the kink method and the MST-inspired approaches suggests that these techniques may provide a useful analytical toolkit for wave and self-force problems in other BH geometries whose radial equations is Heun-like. A notable advantage of the kink method is that it is formulated in terms of a single hypergeometric function (\eqref{psiS1-final},\eqref{Rpsibis} for the In and \eqref{Rpsi_up},\eqref{psiCT0} for the Up), in contrast to the MST formalism, which involves an infinite series of shifted hypergeometric functions. Consequently, directly evaluating the energy loss \eqref{dedt},\eqref{Rlmw} via integration of this closed form hypergeometric expression thereby circumventing the need for integrand expansions presents a promising avenue for future investigation.

All the main results of this paper, including the radiative fluxes, the self-force and the PN-expanded In and Up solutions for $\ell=0,\dots,5$ up to $O(\eta^{10})$, are collected in a Mathematica notebook provided as an ancillary file.


\appendix 

\section{PN expansions of $R_{\rm in}(r)$ and $R_{\rm up}(r)$}\label{appaA}
In this Appendix we report the PN expansions of $R_{\rm in}(r)$ and $R_{\rm up}(r)$ for $\ell=0,1,2$. These expansions are obtained both from formul{\ae} of the kink method (Subsec. \ref{DCHEwave}) and the MST method (Subsec. \ref{MSTin} and Subsec. \ref{Upsec}).
For the first we used \eqref{psiS1-final}-\eqref{psiCT0}, while for the second we used \eqref{RinMST} and \eqref{RupMST}.\\
The expansions from these two methods are in perfect agreement and the common results are listed in the following: 
\begin{widetext}
    \begin{align}
        R_{\rm in}^{(\ell=0)}(r)&=\frac{1}{2} \left(3
   \sqrt{5}-5\right)+\eta^2\Big[\frac{1}{12} \left(5-3
   \sqrt{5}\right) r^2 \omega ^2\Big]+\eta^3\Big[-\frac{10 i \left(7+3 \sqrt{5}+2
   \left(9+4 \sqrt{5}\right)
   \gamma \right) Q \omega
   }{105+47 \sqrt{5}}\Big]\nonumber\\
   &+\eta^4\Big[\frac{1}{240} \left(3
   \sqrt{5}-5\right) r \omega ^2
   \left(r^3 \omega ^2-200
   Q\right)\Big]+\eta^5\Big[\frac{5 i Q \omega  \left(6
   \left(9+4 \sqrt{5}\right)
   Q+\left(7+3 \sqrt{5}+2
   \left(9+4 \sqrt{5}\right)
   \gamma \right) r^3 \omega
   ^2\right)}{3 \left(105+47
   \sqrt{5}\right) r}\Big]\nonumber\\
   &+\eta^6\Big[\frac{60 \left(2889+1292
   \sqrt{5}\right) Q^2 \omega ^2
   \log \left(\frac{2 Q^2 \omega
   }{r}\right)}{33825+15127
   \sqrt{5}}-\frac{1}{504 \left(33825+15127
   \sqrt{5}\right)}\Big(168
   \left(-103304-46199
   \sqrt{5}\right.\nonumber\\
   &\left.-60 \left(6460+2889
   \sqrt{5}\right) \gamma +60
   \left(2889+1292
   \sqrt{5}\right) \gamma ^2-10
   \left(2889+1292
   \sqrt{5}\right) \pi ^2\right)
   Q^2 \omega ^2\nonumber\\
   &-952
   \left(2889+1292
   \sqrt{5}\right) Q r^3 \omega
   ^4+\left(2889+1292
   \sqrt{5}\right) r^6 \omega
   ^6\Big)\Big]+O\left(\eta^7\right)\nonumber\\
   R_{\rm up}^{(\ell=0)}(r)&=\frac{i(15+5\sqrt{5})}{(30+14\sqrt{5})r\omega}+\frac{\eta}{4}(5-3\sqrt{5})-\eta^2\frac{5 i
   \left(3+\sqrt{5}\right)
   \left(r^3 \omega ^2-2
   Q\right)}{4 \left(15+7
   \sqrt{5}\right) r^2 \omega }+\eta^3\frac{60
   \left(2+\left(3+\sqrt{5}\right) \gamma \right) Q+5
   \left(3+\sqrt{5}\right) r^3
   \omega ^2}{12 \left(15+7
   \sqrt{5}\right) r}\nonumber\\
   &+\eta^4\Big[\frac{5 i \left(24
   \left(3+\sqrt{5}\right)
   Q^2-12 \left(-35-9 \sqrt{5}+4
   \left(3+\sqrt{5}\right)
   \gamma -4 i
   \left(3+\sqrt{5}\right) \pi
   \right) Q r^3 \omega
   ^2+\left(3+\sqrt{5}\right)
   r^6 \omega ^4\right)}{48
   \left(15+7 \sqrt{5}\right)
   r^3 \omega }\nonumber\\
   &-\frac{10 i
   \left(3+\sqrt{5}\right) Q
   \omega  \log (2 r \omega
   )}{15+7 \sqrt{5}}\Big]-\frac{\eta^5}{48 \left(15+7
   \sqrt{5}\right) r^2}\Big[-240
   \left(2+\left(3+\sqrt{5}\right) \gamma \right) Q^2\nonumber\\
   &+40
   \left(-9-5 \sqrt{5}+3
   \left(3+\sqrt{5}\right)
   \gamma \right) Q r^3 \omega
   ^2+\left(3+\sqrt{5}\right)
   r^6 \omega ^4\Big]+\eta^6\Big[\frac{20 i \left(20+9
   \sqrt{5}\right) Q \omega 
   \left(9 Q+r^3 \omega
   ^2\right) \log (2 r \omega
   )}{3 \left(15+7
   \sqrt{5}\right)^2 r}\nonumber\\
   &+\frac{1}{72 \left(15+7
   \sqrt{5}\right)^2 r^4 \omega
   }\Big(720 i \left(20+9
   \sqrt{5}\right) Q^3+24
   \left(i \left(2055+889
   \sqrt{5}\right)+300 i
   \left(9+4 \sqrt{5}\right)
   \gamma -60 i \left(20+9
   \sqrt{5}\right) \gamma ^2\right.\nonumber\\
   &\left.{+}30
   \left(105{+}47 \sqrt{5}\right)
   \pi {+}10 i \left(20{+}9
   \sqrt{5}\right) \pi ^2\right)
   Q^2 r^3 \omega ^2{+}10 \left({-}i
   \left(1640{+}741
   \sqrt{5}\right){+}24 i
   \left(20{+}9 \sqrt{5}\right)
   \gamma \right.\nonumber\\
   &\left.{+}24 \left(20{+}9
   \sqrt{5}\right) \pi \right) Q
   r^6 \omega ^4{-}i \left(20{+}9
   \sqrt{5}\right) r^9 \omega
   ^6\Big)\Big]+O\left(\eta^7\right)\nonumber\\
R_{\rm in}^{(\ell=1)}(r)&=2r+\eta^2\Big[-2 Q-\frac{1}{5} r^3 \omega ^2\Big]-4 i\eta^3 (\gamma -1) Q r \omega+\eta^4\Big[\frac{r^5 \omega
   ^4}{140}-\frac{7}{5} Q r^2
   \omega ^2\Big]+\eta^5\Big[\frac{2}{5} i (\gamma -1) Q
   \omega  \left(10 Q+r^3 \omega
   ^2\right)\Big]\nonumber\\
   &-\eta^6\frac{r \omega ^2 \left(-30240
   Q^2 \log \left(\frac{2 Q^2
   \omega }{r}\right)+504
   \left(189+60 (\gamma -3)
   \gamma -10 \pi ^2\right)
   Q^2-906 Q r^3 \omega ^2+r^6
   \omega ^4\right)}{7560}+O\left(\eta^7\right)\nonumber\\
R_{\rm up}^{(\ell=1)}(r)&=-\frac{i}{4 r^2 \omega ^3}-\eta^2\frac{i \left(4 Q+r^3 \omega
   ^2\right)}{8 r^3 \omega ^3} +\eta^3\Big[\frac{r}{12}-\frac{(\gamma -1)
   Q}{2 r^2 \omega ^2}\Big]+\eta^4\Big[\frac{i \left(-24 Q^2-16 Q r^3
   \omega ^2+r^6 \omega
   ^4\right)}{32 r^4 \omega ^3}\Big]\nonumber\\
   &+\eta^5\Big[-\frac{(\gamma -1) Q^2}{r^3
   \omega
   ^2}+\left(\frac{1}{6}-\frac{\gamma }{4}\right)
   Q-\frac{1}{120} r^3 \omega ^2\Big]+\eta^6\Big[\left(\frac{1}{3} i Q r \omega
   -\frac{i Q^2}{2 r^2 \omega
   }\right) \log (2 r \omega
   )\nonumber\\
   &+\frac{i \left(-576 Q^3+48
   (18+6 (\gamma -3) \gamma -\pi
    (\pi -6 i)) Q^2 r^3 \omega
   ^2+4 (-61+24 \gamma -24 i \pi
   ) Q r^6 \omega ^4-r^9\omega^6\right)}{576 r^5
   \omega ^3}\Big]+O\left(\eta^7\right)\nonumber \\
R_{\rm in}^{(\ell=2)}(r) &= 12 r^2 - \frac{6}{7} \eta^2 \left(28 Q r + r^4 \omega^2\right) - 12 i \eta^3 (-3 + 2\gamma) Q r^2 \omega \notag \\
&\quad+ \eta^4 \left(12 Q^2 - \frac{32}{7} Q r^3 \omega^2 + \frac{1}{42} r^6 \omega^4\right) + \frac{6}{7} i \eta^5 (-3 + 2\gamma) Q r \omega \left(28 Q + r^3 \omega^2\right)  + \notag \\
& \quad +\frac{1}{13860} \eta^6 r^2 \omega^2 \Big[ -1584 \Big(906 + 42\gamma(-18 + 5\gamma) - 35\pi^2\Big) Q^2  + 4620 Q r^3 \omega^2 - 5 r^6 \omega^4 + 199584 Q^2 \ln\left(\frac{2 Q^2 \omega}{r}\right) \Big]  - \notag \\
&\quad -\frac{1}{42} i \eta^7 (-3 + 2\gamma) Q \omega \left(504 Q^2 - 192 Q r^3 \omega^2 + r^6 \omega^4\right) + \notag \\
&\quad +\eta^8 \Big[ \frac{4}{35} \Big(2193 + 84\gamma(-18 + 5\gamma) - 70\pi^2\Big) Q^3 r \omega^2  + \frac{1}{1470} \Big(12805 + 504\gamma(-18 + 5\gamma) - 420\pi^2\Big) Q^2 r^4 \omega^4 -\notag \\
&\quad - \frac{148}{17325} Q r^7 \omega^6 + \frac{1}{288288} r^{10} \omega^8  - \frac{36}{35} Q^2 r \omega^2 \left(28 Q + r^3 \omega^2\right) \ln\left(\frac{2 Q^2 \omega}{r}\right) \Big] + \mathcal{O}\left(\eta^9\right)
\label{RinL2}
\end{align}

    \section{PN solutions}\label{appaB}
    \bea
R_{\rm in,PN}^{\ell}(r)&=&r^{\ell }-\eta^2\frac{r^{\ell -1} \left(4 Q \ell
   ^2+6 Q \ell +r^3 \omega
   ^2\right)}{2 (2 \ell +3)}+\frac{\eta^4}{8 (\ell +1)
   (2 \ell +3) (2 \ell +5)}\Bigg[r^{\ell -2} \left(\ell 
   \left(-60 Q^2-180 Q r^3 \omega
   ^2+r^6 \omega ^4\right)\right.\nonumber\\
   &+&\left.16 Q^2
   \ell ^5+64 Q^2 \ell ^4-200 Q
   r^3 \omega ^2+4 Q \ell ^3
   \left(11 Q+2 r^3 \omega
   ^2\right)-4 Q \ell ^2 \left(16
   Q+5 r^3 \omega ^2\right)+r^6
   \omega ^4\right)\Bigg]\nonumber\\
   &+&\eta^6\Bigg\{-\frac{1}{48
   (\ell +1) (\ell +2) (2 \ell
   +1)^2 (2 \ell +3) (2 \ell +5)
   (2 \ell +7)}\Bigg[r^{\ell -3} \left(-48 \ell
   ^6 \left(565 Q^3-Q r^6 \omega
   ^4\right)+256 Q^3 \ell
   ^{10}\right.\nonumber\\
   &+&\left.2176 Q^3 \ell ^9+5248
   Q^3 \ell ^8-2496 Q^3 \ell
   ^7-60480 Q^2 r^3 \omega ^2+288
   Q^2 r^3 \omega ^2 \left(16
   \ell ^6+176 \ell ^5+760 \ell
   ^4+1640 \ell ^3+1849 \ell
   ^2\right.\right.\nonumber\\
   &{+}&\left.\left.1019 \ell {+}210\right) \log
   (r){-}24 Q \ell ^5 \left(1349
   Q^2{+}96 Q r^3 \omega ^2{+}9 r^6
   \omega ^4\right){+}4 \ell ^4
   \left(908 Q^3{-}6048 Q^2 r^3
   \omega ^2{-}1009 Q r^6 \omega
   ^4{+}r^9 \omega ^6\right)\right.\nonumber\\
   &+&\left.2 \ell
   ^3 \left(14668 Q^3{-}48672 Q^2
   r^3 \omega ^2{-}6989 Q r^6
   \omega ^4{+}8 r^9 \omega
   ^6\right){+}\ell ^2 \left(17984
   Q^3{-}187488 Q^2 r^3 \omega
   ^2{-}18690 Q r^6 \omega ^4{+}21
   r^9 \omega ^6\right)\right.\nonumber\\
   &+&\left.\ell 
   \left(3360 Q^3-172512 Q^2 r^3
   \omega ^2-10148 Q r^6 \omega
   ^4+11 r^9 \omega
   ^6\right)-1904 Q r^6 \omega
   ^4+2 r^9 \omega ^6\right)\Bigg]\Bigg\}+O\left(\eta^7\right)\nonumber\\
R_{\rm up,PN}^{\ell}(r)&=& R_{\rm in,PN}^{-\ell-1}(r)
    \eea
\end{widetext}
    \section{Computation of $P^{(n)}_0(y)$, $P^{(n)}_1(y)$ and $P^{(n)}_j(y)$  }\label{Ricc}
    In this Appendix we want to show that the equation
    \bea
&2\Pi_<^{(0)}(y) \Pi_{<}^{(n)}(y) + \frac{d}{dy} \Pi_{<}^{(n)}(y) \nonumber \\
&+ \sum\limits_{m=1}^{n-1} \Pi_{<}^{(m)}(y) \Pi_{<}^{(n-m)}(y)  = V^{(n)}_{<}(y)
\label{recursionPi}
\eea
admits as solution the 
\begin{equation}
   \Pi _{<}^{(n)}(y) = \frac{d}{dy}\left[P^{(n)}_0(y)+  \sum_{j=1}^nP_{j}^{(n)}(y)  \frac{d^{j-1}}{dy^{j-1}}\Pi_{<}^{(0)}(y) \right ]\,.
   \label{Pin2}
\end{equation} 
In general using \eqref{Ricc0} it is clear that the $n$-th derivative of $ \Pi_<^{(0)} $ can be written in terms of its powers as 
\begin{align}
\frac{d^{n} \Pi_<^{(0)} (y)}{dy^{n}} &= a_{n,n+1} \left[ \Pi_<^{(0)}(y) \right]^{n+1} + \sum_{k=0}^{n-1} a_{n, k}  \left[ \Pi_<^{(0)}(y) \right]^{k}
\label{derPin}
\end{align}
where the coefficients $a_{n,k}$ can be computed using the following recursion equations
\bea
&a_{n+1,j}=a'_{n,j}-(j-1) a_{n,j-1} + (j+1) V^{(0)}_{<}(y)a_{n,j+1} \nonumber \\
&   j=0,1,2,\dots, n+2
\eea
with initial conditions $a_{1,0}=V^{(0)}_{<}(y)$ and $a_{1,2}=-1$. Using this we can finally write \eqref{recursionPi} as
\begin{align}
        &\mathcal{A}_{n,n+2} [\Pi_<^{(0)}]^{n+2} +    \mathcal{A}_{n,n+1} [\Pi_<^{(0)}]^{n+1}+    \mathcal{A}_{n,n} [\Pi_<^{(0)}]^{n}+ \dots  +  \nonumber \\
        &+\mathcal{A}_{n,1} [\Pi_<^{(0)}]+  \mathcal{A}_{n,0} =V^{(n)}_<(y) 
    \end{align}
    At this point we can set  $\mathcal{A}_{n,m}=0$  and $\mathcal{A}_{n,0} =V^{(n)}_<(y) $ to find some relations between the various $P^{(n)}_{m}$. \\
    For example for $n=1$ we find
    \bea
    P^{(1)}_0(y)= - \frac{ 1}{2}\frac{d P^{(1)}_1(y)}{dy} \label{P10}
    \eea
    and 
    \begin{equation}
    -\frac{1}{2} P^{(1)'''}_1(y)+2 V^{(0)}_{<}(y) P^{(1)'}_1(y) + V^{(0)'}_{<}(y) P^{(1)}_1(y)=V^{(1)}_<(y)
    \label{eqP11}
\end{equation}
with polynomial solution
\beq
P^{(1)}_{1}(y) = \frac{4 q_{1}}{4k^{2}-1} e^{y}
\eeq
For $n=2$ we find
\bea
&P_{2}^{(2)}(y)&=\frac{P_{1}^{(1)}(y)^2}{2!} \label{P22} \\
&P^{(2)}_{0}(y) &= \frac{1}{2} \left( P^{(1)'}_{1}(y)^{2} - 2 P^{(2)'}_{1}(y) \right) \label{P20}
\eea
and 
\begin{align}
      &-\frac{1}{2} P^{(2)'''}_1(y)+2 V^{(0)}_{<}(y) P^{(2)'}_1(y) + V^{(0)'}_{<}(y) P^{(2)}_1(y) =\nonumber \\
      &= V^{(2)}_<(y) -F_2(y) 
    \label{eqP21}
\end{align}
with 
\begin{align}
&F_2(y)=\frac{1}{2}P_{1}^{(1)}(y)^2V^{(0)''}_{<}(y)+P_{1}^{(1)}(y)'^2V^{(0)}_{<}(y)+\nonumber \\
&+2P_{1}^{(1)}(y)P_{1}^{(1)'}(y)V^{(0)'}_{<}(y)+\frac{3}{4}P_{1}^{(1)''}(y)^2+\frac{1}{2}P_{1}^{(1)'}(y)P_{1}^{(1)'''}(y)
\end{align}
with polynomial solution
\bea
P^{(2)}_{1}(y)&= \frac{\left(-3 - 48k^4 + 28q_1^2 + 8k^2(3 + 10q_1^2)\right) q_2}{(-1 + k^2)(-1 + 4k^2)^3} e^{y}+ \nonumber \\
&+\frac{1 + 16k^4 - 20q_1^2 - 8k^2(1 + 2q_1^2)}{4(1 - 4k^2)^2(-1 + k^2)} e^{2y}
\eea
For $n=3$ we find 
{\footnotesize
\bea
&P^{(3)}_{3}(y) &=\frac{P_{1}^{(1)}(y)^{3}}{3!} \nonumber \\
&P^{(3)}_{2}(y) &= P_{1}^{(1)}(y)P_{1}^{(2)}(y) \label{P3} \\
&P^{(3)}_{0}(y) &=\frac{1}{6} \left( -\left( {P_1^{(1)}}'(y) \right)^3 + 3 {P_1^{(1)}}'(y) {P_1^{(2)}}'(y) - 3 {P_1^{(3)}}'(y) \right)\nonumber 
\eea
}where $P^{(3)}_{1}(y)$ satisfies the differential equation 
\begin{align}
      & -\frac{1}{2} P^{(3)'''}_1(y)+2 V^{(0)}_{<}(y) P^{(3)'}_1(y) + V^{(0)'}_{<}(y) P^{(3)}_1(y) =\nonumber \\
      &= V^{(3)}_<(y) -F_3(y) 
    \label{eqP31}
\end{align}
where $F_3(y)$ is computed from the previously computed $P_{1}^{(2)}(y),P_{1}^{(1)}(y)$ and  $V^{(0)}_{<}(y)$. 
Following the same strategy as the in cases $n=1,2,3$ we can compute recursively the $P_{1}^{(n)}(y)$ and then reconstruct all the other polynomials $P_{j}^{(n)}(y)$ and $P_{0}^{(n)}(y)$, like in \eqref{P10}, \eqref{P22}, \eqref{P20}, \eqref{P3}. 
The explicit results obtained up to $n=4$ suggest the following combinatorial relations between the polynomials $P_{m}^{(n)}(y)$ and $P_{1}^{(l)}(y)$ with $l \leq n$:
\begin{widetext}
\beq
P_{m}^{(n)}(y) := \frac{1}{m!} \hat{B}_{n,m}\left( P_{1}^{(1)}(y) , P_{1}^{(2)}(y), \dots, P_{1}^{(n-m+1)}(y) \right), \qquad 1 \leq m \leq n \,,
    \label{PnkBell-ord1}
\eeq
\beq
 P^{(n)}_{0}(y) := -\frac{1}{2} \sum_{m=1}^{n} \frac{(-1)^{m+1}} {m} \hat{B}_{n,m}\left(P^{(1)\prime}_{1}(y),P^{(2)\prime}_{1}(y), \dots, P^{(n-m+1)\prime}_{1}(y) \right) \,.
\label{Pn01}
\eeq
\end{widetext}
In \eqref{PnkBell-ord1}-\eqref{Pn01} the $ \hat{B}_{n,m}$ are the so-called ordinary Bell polynomials \cite{comtet74}.
 \section{Riccati solution for $\ell=0$  }\label{FloquetL0}
If we look for the general ($ 2 \pi i$ periodic) solution to the Riccati equation \eqref{eq:riccati} in the series form \eqref{PiwQ}, we have to satisfy at leading order 
\begin{align}
    &(\Pi_0^{(\ell=0)})'(y)+(\Pi_0^{(\ell=0)}(y))^2=\frac{1}{4}
    \label{eq:leading} 
\end{align}
which admits the general $ 2 \pi i$ periodic solution
\begin{align}
&\Pi_0^{(\ell=0)}(y)=\frac{1}{2} \left( \frac{e^{y}-e^{y_{0}}}{e^{y}+e^{y_{0}}} \right),
\end{align}
where \(y_0\) is an integration constant. At order \((\omega Q)^2\) we have
\begin{equation}
    (\Pi_2^{(\ell=0)})'(y)+2\Pi_0^{(\ell=0)}(y)\Pi_2^{(\ell=0)}(y)=-W(y)
    \label{eq:pi2eq}
\end{equation}
with solution
\begin{equation}
    \Pi_2^{(\ell=0)}(y)=\frac{1}{I(y)}
    \left[
       C_2-\int^y_{y_{0}} I(s)W(s)\,ds
    \right],
    \label{eq:pi2solution}
\end{equation}
where \(C_2\) is an integration constant and 
\begin{align}
    I(y)=\exp\left(2\int_{y_{0}}^y\Pi_0^{(\ell=0)}(s)\,ds\right)=\frac{(e^{y} +e^{y_{0}})^{2}}{4 e^{y+y_{0}}}
\end{align}
Since \(I(y)\) is \(2\pi i\)-periodic, a necessary
and sufficient condition for the solution \(\Pi_2^{(\ell=0)}(y)\) to be
\(2\pi i\)-periodic is
\begin{equation}
    \int_y^{y+2\pi i}I(s)W(s)\,ds=0 \, , 
    \label{solvability}
\end{equation}
which gives for the integration constant $y_{0}$ the two values
\begin{equation}
   e^{y_0}=\frac{-3\pm\sqrt{5}}{2}\,,
\end{equation}
that identify the two different Floquet solutions.\\
At general order  \((\omega Q)^{2n}\)  we have
\beq
\Pi_{2n}^{(\ell=0)'}(y) + 2\Pi^{(\ell=0)}_0(y)\Pi_{2n}^{(\ell=0)}(y) = G_{2n}(y),
\label{generalPin}
\eeq
where the first driving terms read 
\begin{align} 
G_2(y) &= -W(y), \\ 
G_4(y) &= -\Pi_2(y)^2, \\ 
G_6(y) &= -2\Pi_2(y)\Pi_4(y). 
\end{align}
Integrating \eqref{generalPin} we would obtain
\beq
\Pi_{2n}^{(\ell=0)}(y) = \frac{1}{I(y)}
    \left[
       C_{2n}+\int^y_{y_{0}} I(s)G_{2n}(s)\,ds
    \right],
    \label{solPin}
\eeq
The $C_{2n}$ is uniquely fixed by imposing the periodicity condition at the next order 
\begin{equation}
    \int_y^{y+2\pi i}I(s)G_{2n+2}(s)\,ds=0.
    \label{solvability2}
\end{equation}
Considering the structure of $G_{2n}(y)$ we can rewrite \eqref{solPin} in the form \eqref{PiL0}.
    \section{$\mathcal{S}$ and $\mathcal{\hat{S}}$}   \label{AppS} 
 \eqref{Sexp} and \eqref{Shexp} for $n=1,2$  are given by
{\footnotesize
\begin{flalign}
\mathcal{S}_{1} &= \frac{2 q_{1}}{1 - 4k^2} e^{y} & \\
\mathcal{S}_{2} &= \frac{8 q_1^2}{(4 k^2 - 1)^2} 
+ \frac{\left[ 3 + 4 q_1^2 + 16 k^4 (3 + 8 q_1^2) - 24 k^2 (1 + 10 q_1^2) \right] q_2}{2 (k^2 - 1)(4 k^2 - 1)^3} \, e^y \nonumber \\
&+ \frac{1 + 4 q_1^2 + k^2 (8 q_1^2 - 4)}{4 (k^2 - 1)(4 k^2 - 1)} \, e^{2y} &
\end{flalign}
}
and
{\footnotesize
\begin{flalign}
\mathcal{\hat{S}}_{1} &= -\frac{4 q_{1}}{1 - 4k^2} e^{y} & \\
\mathcal{\hat{S}}_{2} &=
\frac{\left[ -3 - 48 k^4 + 28 q_1^2 + 8 k^2 (3 + 10 q_1^2) \right] q_2}{(k^2 - 1)(4 k^2 - 1)^3} \, e^y \nonumber \\
&- \frac{1 - 4 k^2 + 12 q_1^2}{4 (k^2 - 1)(4 k^2 - 1)} \, e^{2y} &
\end{flalign}
}

\section*{Acknowledgments}

We would like to thank D. Bini, D. Gregori, A. Geralico, T. Manton, A. Marcian\`o, S. de Jes\'us Quir\'os Araya, C.-H. Shen and A. Tokareva for useful discussions and suggestions. This research was supported in part by the grants GAST (INFN), the Cost actions CaLISTA CA21109 and Theory-Challenges CA22113 "Fundamental Challenges in Theoretical Physics", MSCA CaLIGOLA 2021-SE-01-101086123, the EC Network Gatis and the National Science Foundation under Grant No. NSF PHY-1748958. G.D.R. acknowledges  membership to the Italian Gruppo Nazionale per la Fisica Matematica (GNFM) of the Istituto Nazionale
di Alta Matematica (INDAM). 
GDR would like to thank National Taiwan University (NTU) for its kind hospitality during the completion of part of this work.
FO would like to acknowledge the financial support from the NextGeneration EU through Italian Ministry of University and Research (MUR) under PNRR Mission 4, Component 2, Investment 3.3 (M.D. 630/2024), and also the University of Uppsala for kind hospitality during the completion of this work. GDR, DF, MR thank organisers and participants of the workshop 'Exact techniques and their applications - InTropea26' for the stimulating atmosphere for the development of this work. 

\bibliographystyle{apsrev}
\bibliography{references}

\end{document}